\documentstyle[12pt,epsfig]{article}
\newlength{\dinwidth}
\newlength{\dinmargin}
\newcommand{\ba}{\begin{array}}
\newcommand{\ea}{\end{array}}
\newcommand{\beq}{\begin{equation}}
\newcommand{\eeq}{\end{equation}}
\newcommand{\bea}{\begin{eqnarray}}
\newcommand{\eea}{\end{eqnarray}}

\def\bce{\begin{center}}
\def\ece{\end{center}}

\def\nonu{\nonumber}

\def\pa{\partial}
\def\al{\alpha}

\def\De{\Delta}
\def\ep{\epsilon}

\def\th{\theta}

\def\La{\Lambda}

\def\eps6{{\displaystyle \mathop{\epsilon}^{6}}{}}

\def\nab6{{\displaystyle \mathop{\nabla}^{6}}{}}
\def\to{\rightarrow}

\newcommand{\bean}{\begin{eqnarray*}}
\newcommand{\eean}{\end{eqnarray*}}

\begin{document}

\thispagestyle{empty} \addtocounter{page}{-1}
\begin{flushright}
%NSF-ITP 99-098\\
%KIAS-P03016 \\
TIT-HEP-509 \\
{\tt hep-th/0309156}\\
%revised, Feb., 2002\\
\end{flushright}

\vspace*{1.3cm} 
\centerline{ \Large \bf The Matrix Model Curve Near the 
Singularities }
\vspace*{1.5cm}
\centerline{{\bf Changhyun Ahn}$^1$ and 
{\bf Yutaka Ookouchi}$^2$}
\vspace*{1.0cm} \centerline{\it $^1$Department of Physics,
Kyungpook National University, Taegu 702-701, Korea}
\vspace*{0.2cm} \centerline{\it $^2$Department of Physics,
 Tokyo Institute of
Technology, Tokyo 152-8511, Japan}

\vspace*{0.8cm} \centerline{\tt
ahn@knu.ac.kr,  \qquad 
ookouchi@th.phys.titech.ac.jp}
\vskip2cm

\centerline{\bf Abstract}
\vspace*{0.5cm}

In ${\cal N}=1$ supersymmetric $SO(N)/USp(2N)$ gauge theories with 
the tree-level superpotential $W(\Phi)$
that is an arbitrary polynomial of the adjoint matter 
$\Phi$, the massless fluctuations
about each quantum vacuum are described by $U(1)^n$ gauge theory.
By turning on the parameters of $W(\Phi)$ to the special values,
the singular vacua where the additional fields become massless can be 
reached.    
Using the matrix model prescription, we study 
%the singularities of the $n=0$ branch and 
the intersections of $n=0$ and $n=1$ branches.
The general formula for the matrix model curve at the singularity
which is valid for arbitrary $N$ is 
obtained and this generalizes the previous results for small values of
$N$ from strong-coupling approach.   
Applying the analysis to the degenerated case, we also obtain a 
general matrix model curve which is not only 
valid at a special point but also on the whole branch. 

%The matrix model description of the low energy physics breaks down
%at the singularity.

\baselineskip=18pt
\newpage
\renewcommand{\theequation}{\arabic{section}\mbox{.}\arabic{equation}}

%%%%%%%%%%%%%%%%%%%%%%%%%%%%%%%%%%%%%%%%%%%%%%%%%%%%%%%%%%%%%%%%%%%%%%%%%
\section{Introduction}
\setcounter{equation}{0}
%%%%%%%%%%%%%%%%%%%%%%%%%%%%%%%%%%%%%%%%%%%%%%%%%%%%%%%%%%%%%%%%%%%%%%%%%

\indent

A new recipe for the computation of the exact quantum effective
superpotential for the glueball field
was proposed by Dijkgraaf and Vafa \cite{dv1,dv2,dv3} using 
a zero-dimensional matrix model.   
Extremization of the effective glueball superpotential 
has led to the quantum vacua of the supersymmetric gauge theory.
For ${\cal N}=1$ supersymmetric $U(N)$ gauge theory with the adjoint 
matter $\Phi$, the gauge group $U(N)$ breaks into $\prod_{i=1}^n U(N_i)$
for some $n$. At low energies, the effective theory becomes 
${\cal N}=1$ gauge theory with gauge group $U(1)^n$.
The low energy dynamics have been studied in 
\cite{cdsw,csw,csw1,fer1,fer2,fer3,fo2}.

Recently in \cite{Shih}, the matrix model curve for $U(N)$ gauge theory 
was obtained through the 
glueball approach. For example, the intersections of $n=1$ and $n=2$ 
branches for cubic superpotential occur when a vacuum with gauge group
$U(N_1) \times U(N_2)$ meets a vacuum with gauge group
$U(N)$ where $N=N_1+N_2$. 
The general formulas for the parameters of tree-level superpotential
are functions of $N_1$ and $N_2$. The locations of the $n=1$ and $n=2$
singularities and the expectation value of glueball field at the 
singularities were obtained in \cite{csw} from the ${\cal N}=2$ factorization
problem using the strong-coupling approach \cite{civ,cv}: 
for small values of $N$ it was possible to solve explicitly.
However, this general factorization problem will meet some difficulty
as the $N$ increases. The outcome of \cite{Shih} allows us to write down
the quantum vacua at the singularity for general $(N_1,N_2)$ which will
generalize \cite{csw}.

On the other hand, 
the ${\cal N}=1$ matrix model curve for $SO(N)/USp(2N)$ 
gauge theories
with 
the tree-level superpotential $W(\Phi)$
is characterized by 
\bea
y_m^2 = F_{2(2n+1)}(x) = W_{2n+1}^{\prime}(x)^2+{\cal O}(x^{2n})
\nonu
\eea
where the tree-level superpotential is given by  
\bea
W_{2(n+1)}(\Phi) = \sum_{r=1}^{n+1} \frac{g_{2r}}{2r} \mbox{Tr} \Phi^{2r}.
\label{pot}
\eea
In \cite{ao,afo1,afo2}, the explicit constructions for 
the matrix model curve using the factorization problem
were obtained for small values of $N$, when we 
consider the quartic superpotential ($n=1$), in the strong-coupling
approach. Here the matrix model curve 
possesses an arbitrary parameter. We expect there is a chance to have 
an extra double root  by restricting
ourselves to the particular value for superpotential parameter.   
The singularity arises from an additional monopole becoming 
massless in the strong-coupling description. 
We apply the method of \cite{Shih} to the $SO(N)/USp(2N)$ gauge theories.

In this paper, 
we study how the generic picture can be changed at the strong-coupling 
singularities where the additional fields become massless and the presence
of extra massless fields will lead to an interacting superconformal field
theory. These singularities
can be obtained by turning on the parameters of $W(\Phi)$ to the 
particular values.   
By solving the glueball equations of motion at the $n=0$ and $n=1$
singularity, one gets a general formula for the parameter and fluctuating 
fields at these 
singularities.
Our general formula extends the results of \cite{ao,afo1,afo2} to provide
an information on the ${\cal N}=1$ matrix model curve 
for arbitrary $(N_0, N_1)$. 

In section 2.1, 
in order to find out the glueball equations of motion for given  
effective superpotential, we compute the derivatives of dual periods 
with respect to the fluctuating fields explicitly. This will lead to 
the solutions for the two kinds of parametrization of 
the matrix model curve (\ref{Curve}) 
in terms of $N_0$, $N_1$, and the scale $\La$
of $SO(N)$ gauge theory.  
Based on this general formula, we compare our results with the matrix model
curve from previous results by 
further restricting some parameter of superpotential and we find 
an exact agreement.
In section 2.2, the coupling constant at the singularity goes to vanish
as we compute the derivative of dual period with respect to the glueball
field.

In section 3.1,  
based on the general formula for the matrix model curve of 
$USp(2N)$ gauge theory (\ref{for}), 
we compare our results with the matrix model
curve from strong-coupling approach 
%by 
%further restricting some parameter of superpotential 
and we find 
an exact agreement.
In section 3.2,  the coupling constant corresponding to
the gauge coupling constant of the nontrivial $U(1)$  
at the singularity goes to 
vanish.

In section 4, we apply the method of section 2.1 to the 
degenerated case for $SO(N)$ gauge theory
in which the matrix model curve (\ref{formula2})
is parametrized by two 
variables with one constraint. In this case, we also find an exact 
agreement from strong-couping approach.

%In section 5, we 

In Appendices A and B, we present some detailed computations which are
necessary to sections 2 and 3. 

%%%%%%%%%%%%%%%%%%%%%%%%%%%%%%%%%%%%%%%%%%%%%%%%%%%%%%%%%%%%%%%%%%%%%%%%%%%%%
\section{The $n=0$ and $n=1$ singularity: $SO(N)$ gauge theory}
\setcounter{equation}{0}
%%%%%%%%%%%%%%%%%%%%%%%%%%%%%%%%%%%%%%%%%%%%%%%%%%%%%%%%%%%%%%%%%%%%%%%%%%%%%

%\indent

%One paragraph for the notations: $n$ and so on

%%%%%%%%%%%%%%%%%%%%%%%%%%%%%%%%%%%%%%%%%%%%%%%%%%%%%%%%%%%%%%%%%%%%%%%%%%%%%
\subsection{The glueball equations of motion \label{glueball equations}}
\setcounter{equation}{0}
%%%%%%%%%%%%%%%%%%%%%%%%%%%%%%%%%%%%%%%%%%%%%%%%%%%%%%%%%%%%%%%%%%%%%%%%%%%%%

\indent

Let us consider the intersections of the $n=0$ and $n=1$ branches.
These occur at special values of the tree-level superpotential
parameter and a vacuum  with unbroken 
gauge group \footnote{In several examples we present below, the 
unbroken gauge group contains $U(1)$ factor. We also consider the 
glueball superfield for the $U(1)$ gauge group and extremize
the corresponding glueball field.
Recently 
the issue when glueball superfields 
should be included and extremized
or set to zero has been studied in \cite{ikrsv}.
According to this general prescription 
for how string theory deals with low rank 
gauge groups including $U(1)$ group in the geometric 
dual description, the generalized dual Coxeter number 
for $U(1)$ is 1 which is positive 
and one should include the corresponding glueball superfield
and extremize the glueball 
superpotential with respect to it. 
Therefore, the theory has a dual confining description since
the string theory computes not for the standard gauge theory
but the associated higher rank gauge theory \cite{ikrsv}. }
$SO(N_0) \times U(N_1)$
intersects a vacuum with unbroken gauge group $SO(N)$ with 
\bea
N=N_0 +2N_1.
\nonu
\eea
Although the structure of these singularities has been 
discussed in \cite{ao} implicitly 
by applying the strong-coupling approach,
in this paper we study these intersection singularities in detail using the 
glueball description \cite{Shih}.
We will 
describe the approach to the singularity from the $n=1$
branch  since the approach from the $n=0$ branch generally 
behaves without any singularity: the matrix model curve 
on the $n=0$ branch is regular as we go through the intersection 
with the $n=1$ branch.

Let us take the tree-level superpotential to be  quartic (\ref{pot}).
We expect that the general feature of the analysis for this particular 
superpotential holds for the general superpotential of 
arbitrary degree $2(n+1)$.
As we approach the $n=0$ and $n=1$ singularity, both the 
matrix model curve and the SW curve possess an extra double root.
Using the matrix model curve, one can compute the 
effective glueball superpotential and $U(1)$ gauge coupling 
near the singularity.
The matrix model curve is given by \cite{feng,eot}
\bea
y_m^2 = W_3^{\prime}(x)^2+f_2 x^2 +f_0=
\left(x^2+x_0^2 \right)\left(x^2+x_1^2\right)
\left(x^2+x_2^2\right).
\label{matrixcurve}
\eea
Here we assume that all three branch cuts $[-ix_2, -ix_1],
[-ix_0,ix_0]$, and $[ix_1,ix_2]$ 
are along the 
imaginary axis and the contour of noncompact cycle 
$B_0$ as from the origin to 
the cut-off $\Lambda_0$ is along the real axis. The compact cycles $A_i$ have 
to intersect the noncompact cycles $B_i$ as $(A_i, B_j)=\delta_{ij}$ where
$i=0,1$. 

%
%Since all the branch cuts are along the imaginary axis we 
%can represent matrix model curve for $SO/USp$ gauge theories as follows:
%\begin{eqnarray}
%
%\end{eqnarray}
%where we define $f(x)\equiv f_2x^2+f_0$. 
%
If we parametrize the tree-level superpotential as 
\bea
W_3^{\prime}(x)=
x^3+mx,
\nonu
\eea 
we obtain the following relation,
\bea
m=\frac{1}{2} \left(x_0^2+x_1^2+x_2^2 \right).
\label{constraint}
\eea
The first parametrization of the matrix model curve 
(\ref{matrixcurve}) implies that 
$m$ is a parameter and $f_2$ and $f_0$ are fluctuating fields that
are related to the two glueball fields $S_1$ and $S_0$ respectively.
The second parametrization in terms of the roots $ \pm i x_0, \pm  i x_1$
and $\pm i x_2$ will be more convenient and these fields are subject to
the constraint (\ref{constraint}). 
One can always interchange from one  parametrization to the other 
through the matrix model curve (\ref{matrixcurve}).

We evaluate the derivatives of the dual periods of the matrix model 
curve on the $n=1$ branch, with quartic tree-level superpotential.
The periods $S_i$ of holomorphic 3-form for the deformed geometry 
over compact $A_i$ cycles  and dual periods $\Pi_i$ of holomorphic
3-form  over noncompact
$B_i$ cycles
are written in terms of 
the integrals over $x$-plane ($i=0,1$). 
The periods are the glueball fields $S_i$ 
\footnote{In \cite{fo}, the period 
$S_0$ was an integral over $x$-plane from 
$-\Delta_0$ to $\Delta_0$. 
In our notation here if we replace $x_0\to -i\Delta_0$, 
the $S_i$'s  agree with those in \cite{fo}. 
For the dual period $\Pi_0$, since there is no singularity on 
the Riemann surface $y_m(x)$, we can change the lower limit 
smoothly on the branch cut as $0 \rightarrow \De_0$. 
Taking into account  
this we can see the agreement with \cite{fo}.} \cite{fo,eot}:
\begin{eqnarray}
2 \pi i S_0  & = & \int^{ix_0}_{-ix_0}
y_m \  dx  =  \int^{ix_0}_{-ix_0}
\sqrt{\left(x^2+x_0^2\right)\left(x^2+x_1^2\right)
\left(x^2+x_2^2\right)}\  dx, 
%=
%-\int^{x_0}_{-x_0}\sqrt{(x^2-x_0^2)(x^2-x_1^2)
%(x^2-x_2^2)}  dx, 
\nonumber \\
2\pi i S_1&=& \int^{ix_2}_{ix_1}
y_m \ dx =
\int^{ix_2}_{ix_1}\sqrt{\left(x^2+x_0^2\right)\left(x^2+x_1^2\right)
\left(x^2+x_2^2\right)} \ dx, 
% = -\int_{x_1}^{x_2}\sqrt{(x^2-x_0^2)(x^2-x_1^2)(x^2-x_2^2)} dx,
\nonu
\end{eqnarray}
and their conjugate periods:
\begin{eqnarray}
2\pi i \Pi_0&=&
\int^{\Lambda_0}_{0} y_m \ dx =
\int^{\Lambda_0}_{0}\sqrt{\left(x^2+x_0^2\right)\left(x^2+x_1^2\right)
\left(x^2+x_2^2\right)} \ dx, \nonumber \\
2\pi i \Pi_1&=&
\int^{i\Lambda_0}_{ix_2} y_m  \ dx =
\int^{i\Lambda_0}_{ix_2}\sqrt{\left(x^2+x_0^2\right)\left(x^2+x_1^2\right)
(x^2+x_2^2)} \ dx. 
% = -\int^{\Lambda_0}_{x_2}\sqrt{(x^2-x_0^2)(x^2-x_1^2)(x^2-x_2^2)}
% dx.
\nonu
\end{eqnarray}

These periods provide the effective glueball superpotential \cite{feng}
(See also
\cite{eot,fo,oz,ashoketal,jo,ow}):
\begin{eqnarray}
W_{\mbox{eff}}=  2 \pi i \left[ \left(N_0 - 2 \right) \Pi_0 +2N_1 \Pi_1
\right] -
2 \left(N - 2 \right) S \log 
\left(\frac{\Lambda}{\Lambda_0} \right). 
\label{super}
\end{eqnarray}
One can generalize this to add the $b_1 S_1$ term  but as in \cite{Shih} 
after the trivial calculation $b_1$ must be zero at the 
singularity (according to the computation of Appendix B, $\pa S_1/\pa f_0$
and $\pa S_1/\pa f_2$ are divergent at the singularity and therefore 
in order to have consistent equations of motion $b_1$ should vanish) 
and $S=S_0+2S_1$.
Over a cycle $A_i$ surrounding the $i$-th cut, the periods of
$T(x)$ are \cite{ao}
\bea
N_0= \frac{1}{2\pi i} \oint_{A_0} T(x) dx, \qquad
N_1= \frac{1}{2\pi i} \oint_{A_1} T(x) dx
\label{n0n1}
\eea
where
\bea
T(x) = \frac{d}{d x} \log \left( P_{N}(x) + \sqrt{P_N^2(x)- 4 
x^{2(1+\ep)} \La^{2N-2(1+\ep)}} \right), \qquad
P_N(x)= \mbox{det} \left( x -\Phi \right)
\nonu
\eea
with $\ep =0$ for  $N$ odd, and $\ep =1$ for  $N$ even.
Let us stress that $N_0$ and $N_1$ (they are always integers) 
are defined as the on-shell periods 
of the one-form $T(x)$ 
at the $n=0$ and $n=1$ singularity. As we will see below,
the  precise values of $(N_0,N_1)$ can be determined
through (\ref{n0n1}) with an appropriate choice of cycles.

One can represent a general formula for the derivatives of
the effective superpotential, which holds for the general $(N_0,N_1)$
with the help of Appendix A. 
As we mentioned before, $x_i (i=0,1,2)$ lie on the imaginary line and
$x_0< x_1 < x_2$ which allows us to compute the elliptic integrals
without any ambiguities.
In principle, this formula provides the solutions for the
$x_i$ in terms of the parameters $N_0, N_1$, and $m$. Moreover, 
according to the matching conditions (\ref{matrixcurve}) 
between the two parametrizations,
this will lead to the expectation values of the fields $f_2$ and $f_0$ 
eventually.   
Although our general formula is valid not just near the $n=0$
and $n=1$ singularity, we are interested in the solutions near the
$n=0$ and $n=1$ singularity \footnote{
Other kind of singularities in different context may arise.
For example, ${\cal N}=1$ Argyres-Douglas (AD) points.
See the paper \cite{es}.
It would be interesting to study the effective superpotential
in both the glueball and the strong-coupling approach 
\cite{gan}.   }.

To compute the expectation values on the $n=0$ and $n=1$ singularity, 
we can simply take a limit $x_1\to x_0$. Or one can compute those 
by starting from the general formula and then substituting the 
condition 
$x_1=x_0$ at the final stage. We will elaborate this in  Appendix A.

%%%%%%%%%%%%%%%%%%%%%%%%%%%%%%%%%%%%%%%%%%%%%%
$\bullet$ The equation of motion for  a field $f_0$
%: 
%$\frac{\pa W_{\mbox{eff}} }{ \pa f_0}=0$
%%%%%%%%%%%%%%%%%%%%%%%%%%%%%%%%%%%%%%%%%%%%%%%%%

The derivative of $\Pi_0$ with respect to $f_0$ is given, by recognizing 
$f_0 =x_0^4 x_2^2$, the $x$-independent part inside of the square root
in the dual periods, 
as 
%(we make a change of variable $t=x^2$)
\begin{eqnarray}
4\pi i \frac{\partial \Pi_0}{\partial f_0}&=&
%\int_{0}^{\Lambda_0}
%\frac{1}{\sqrt{(x^2+x_0^2)^2(x^2+x_2^2)}}\ dx 
%= 
\int_{0}^{\Lambda_0^2}\frac{dt}{2\left(t+x_0^2\right)\sqrt{t
\left(t+x_2^2\right)}} 
\nonumber \\
&=& \frac{1}{2x_0 \sqrt{x^2_2-x_0^2}} \left[ \sin^{-1} \left(
\frac{\left(x_2^2-2x_0^2\right)\left(t+x_0^2\right)+2x_0^2
\left(x_0^2-x_2^2\right)}
{x_2^2\left(t+x_0^2\right)} \right) \right]^{\Lambda_0^2}_{0} \nonumber \\
& \simeq &\frac{1}{2x_0 \sqrt{x^2_2-x_0^2}} \left[\sin^{-1}
\left(\frac{x_2^2-2x_0^2}{x_2^2} \right)+\frac{\pi}{2} \right].
\label{del00}
\end{eqnarray}
where we change the integration variable as  $t=x^2$ and in the last
expression we take $\La_0$ very large. 

Similarly, one can execute the 
integral (we change a variable $t=-x^2$)
\begin{eqnarray}
4\pi i \frac{\partial \Pi_1}{\partial f_0}&=&
%-\int_{x_2}^{\Lambda_0}
%\frac{1}{\sqrt{(y^2-x_0^2)^2(y^2-x_2^2)}}\ dy 
%= 
-\int_{x_2^2}^{\Lambda_0^2}\frac{dt}{2\left(t-x_0^2\right)
\sqrt{t\left(t-x_2^2\right)}} 
\nonumber \\
&=& \frac{-1}{2x_0 \sqrt{x_2^2-x_0^2}}\left[ \sin^{-1} \left(
\frac{\left(-x_2^2+2x_0^2\right)\left(t-x_0^2\right)+
2x_0^2\left(x_0^2-x_2^2\right))}
{x_2^2\left(t-x_0^2\right)}\right) \right]^{\Lambda_0^2}_{x_2^2} \nonumber \\
& \simeq & \frac{-1}{2x_0 \sqrt{x_2^2-x_0^2}} \left[\sin^{-1}
\left( \frac{-x_2^2+2x_0^2}{x_2^2} \right)+\frac{\pi}{2} \right].
\label{del10}
\end{eqnarray}
We also drop the irrelevant terms in the last equation as we take the 
large limit of $\La_0$.

By using these two results (\ref{del00}) and (\ref{del10}) and taking 
$\partial W_{\mbox{eff}}/\partial f_0=0$ ($W_{\mbox{eff}}$ is given by
(\ref{super})),
we obtain one equation of motion for $f_0$ 
after manipulating the trigonometric
functions: 
\begin{eqnarray}
(N_0-2)\frac{\partial \Pi_0}{\partial f_0}+2N_1 
\frac{\partial \Pi_1}{\partial f_0}=0 \iff -
\left( \frac{x_2^2-2x_0^2}
{x_2^2} \right) =\cos \left( \frac{2\pi N_1}{N-2} \right).
\label{solution1}
\end{eqnarray}

%%%%%%%%%%%%%%%%%%%%%%%%%%%%%%%%%%%%%%%%%%%%%%
$\bullet$ The equation of motion for a field $f_2$
%: 
%$\frac{\pa W_{\mbox{eff}} }{ \pa f_2}=0$
%%%%%%%%%%%%%%%%%%%%%%%%%%%%%%%%%%%%%%%%%%%%%%%

The derivative of $\Pi_0$ with respect to $f_2$ is given, 
by recognizing 
%$f_2$, 
the $x^2$ part inside of the square root in the dual periods, 
as (we change a variable $t=x^2$)
\begin{eqnarray}
4\pi i \frac{\partial \Pi_0}{\partial f_2}&=&
%\int_{0}^
%{\Lambda_0}\frac{x^2}{\sqrt{(x^2+x_0^2)^2(x^2+x_2^2)}}\ dx  
%= 
\int_{0}^{\Lambda_0^2}\frac{\sqrt{t}}{2\left(t+x_0^2\right)
\sqrt{\left(t+x_2^2\right)}}
\ dt 
= \left[\frac{1}{2} \log \left( 2t+x_2^2+
2\sqrt{t\left(t+x_2^2\right)} \right) 
\right]^{\Lambda^2_0}_{0}-4\pi i
 x_0^2 \frac{\partial \Pi_0}{\partial f_0} \nonumber \\
&\simeq & \frac{1}{2}
\log \Bigg|  \frac{4\Lambda_0^2}{x_2^2} \Bigg| -4\pi i x_0^2 
\frac{\partial \Pi_0}{\partial f_0}.
\label{02der}
\end{eqnarray}
Note that the second term was given by (\ref{del00}) multiplied by $x_0^2$.
Moreover 
one obtains  by using a change of variable $t=-x^2$
\begin{eqnarray}
4\pi i \frac{\partial \Pi_1}{\partial f_2}&=&
%\int_{x_2}^{\Lambda_0}\frac{y^2}{\sqrt{(y^2-x_0^2)^2(y^2-x_2^2)}}
%\ dy 
%= 
\int_{x_2^2}^{\Lambda_0^2}\frac{t}{2\left(t-x_0^2\right)\sqrt{t
\left(t-x_2^2\right)}}
\ dt
= \frac{1}{2} \left[\log \left( 2t-x_2^2+2\sqrt{t\left(t-x_2^2\right)} \right) 
\right]^{\Lambda^2_0}_{x_2^2}-4\pi i x_0^2 \frac{\partial 
\Pi_1}{\partial f_0} \nonumber \\
&\simeq & \frac{1}{2}
\log \Bigg| \frac{4\Lambda_0^2}{x_2^2} \Bigg| -4\pi i x_0^2 
\frac{\partial \Pi_1}{\partial f_0}
\label{12der}
\end{eqnarray}
where the second term is given by (\ref{del10}) multiplied by $x_0^2$.

By using these two results (\ref{02der}) and (\ref{12der}) and taking
$\partial W_{\mbox{eff}}/\partial f_2=0$,
we obtain one equation of motion for $f_2$ by cooperating with the 
equation of motion for $f_0$ (\ref{solution1}): 
\begin{eqnarray}
(N_0-2)\frac{\partial \Pi_0}{\partial f_2}+2N_1 \frac{\partial 
\Pi_1}{\partial f_2} + (N-2) \log \left( \frac{\La}{\La_0}\right)=
0 \iff  \frac{1}{2} \left(N-2 \right)\log \Bigg| 
\frac{x_2^2}{4\Lambda^2} \Bigg| =0
\label{solution2}
\end{eqnarray}
which implies the $(N-2)$ branches labeled by $\eta$.
That is, $\eta$ is the $(N-2)$-th root of unity for $N$ even and
the $(N-2)$-th root of minus unity. That is,
\bea
\eta^{N-2} =1, \;\; \mbox{for} \;\; N \;\; \mbox{even}, \qquad
\eta^{N-2} =-1 \;\; \mbox{for} \;\; N \;\; \mbox{odd}.
\nonu
\eea

From the two solutions (\ref{solution1}) and (\ref{solution2}), 
we obtain the following results together with (\ref{constraint}),
\bea
x_2^2&=&4\eta \Lambda^2,\qquad x_0^2=2\eta \Lambda^2 
\left( 1+\cos \frac{2\pi N_1}{N-2} \right),
\qquad
m  =  2\eta \Lambda^2 
\left( 2+\cos \frac{2\pi N_1}{N-2} \right).
\label{sol}
\eea
Matching the two parametrizations in (\ref{matrixcurve}),
we obtain the expectation values of the fields $f_2$ and $f_0$
at the double root singularity:
\bea
\langle f_2 \rangle &=& 
%-\eta^2 \left(x_0^2+\frac{x_2^2}{2} \right)^2+
%x_0^4+2x_2^2x_0^2 
 4 \eta^2 \Lambda^4 \left(1+2\cos \frac{2\pi N_1}{N-2} 
\right),
\nonumber \\
\langle f_0 \rangle & = & 
16 \eta^3 \La^6 \left( 1+\cos \frac{2\pi N_1}{N-2} \right)^2.  
\label{sol1}
\eea 
Since the total glueball field is related to $f_2$ through 
$S=-\frac{f_2}{4}$, we also obtain
the 
expectation value of the glueball field at the $n=0$ and $n=1$
singularity:
\bea
\langle S \rangle &=&- \eta^2 
\Lambda^4 \left(1+2\cos \frac{2\pi N_1}{N-2} 
\right). 
\label{formula}
\end{eqnarray}
The general formulas (\ref{sol}), (\ref{sol1}) and (\ref{formula})
for the matrix model curve, the parameter of the tree level 
superpotential and the expectation value of the glueball field 
are new. Previously, the locations of $n=0$ and $n=1$ singularities
and the expectation value of the glueball field  can be obtained
only for small numbers  of $N$ where the factorization problem could be 
solved explicitly \cite{ao} by restricting the superpotential parameter
further.
The difficulty of the solving the general factorization problem when
$N$ is large can be avoided by looking at both 
the glueball equations of motion 
at the $n=0$ and $n=1$ singularity \cite{Shih}
and the quantum vacua at the singularity
for general $(N_0, N_1)$.
Now the matrix model curve can be summarized as
\bea
y_m^2 & = & x^2 \left[ x^2+      
 2 \eta \La^2  \left( 2+ c \right) \right]^2 +
 4 \eta^2 \Lambda^4 \left(1+2 c 
\right) x^2 
 + 
16 \eta^3 \La^6 \left( 1+ c \right)^2 
\nonu \\  
&= & \left[x^2 + 2\eta \Lambda^2 
\left( 1+ c \right) \right]^2 
\left( x^2 + 4\eta \Lambda^2  \right), \qquad c \equiv 
\cos \left( \frac{2\pi N_1}{N-2} \right) 
\label{Curve}
\eea
where $N_0$ and $N_1$ are given in (\ref{n0n1}) together with $N=N_0+2N_1$. 

We can explicitly demonstrate 
this general result by comparing them with the results 
obtained in \cite{ao}. Let us consider $SO(N)$ case 
where $N=4,5,6,7$, and 8. 

%%%%%%%%%%%%%%%%%%%%%%
$\bullet$ $SO(4)$
%%%%%%%%%%%%%%%%%%%%%%

For $SO(4)$ case, we only consider the
breaking pattern $SO(4)\to SO(2)\times U(1)$. In this case from 
the solutions (\ref{sol}), (\ref{sol1}) and (\ref{formula}), 
one predicts 
the matrix model curve, by putting $(N,N_1)=(4,1)$ 
(note that $N=N_0+2N_1$), in terms of two parametrizations
\bea
y_m^2 = x^2 \left( x^2+ m \right)^2 - 4  \La^4 x^2
= x^4 \left( x^2 + 4 \eta \La^2 \right)
\nonu
\eea
with $m = 2 \eta \La^2$ where $\eta$ is 2-nd root of unity 
and also we find $\langle S \rangle=
\La^4$. In \cite{ao}, the factorization problem 
resulted in the matrix model curve 
$
\widetilde{y_m}^2 = x^2 \left( x^2- v^2 \right)^2 - 4  \La^4 x^2
$ and one can easily check that the intersections with the $n=0$ branch 
occur at $v^2 = -2 \eta \La^2$ where $\widetilde{y_m}^2$ has an additional 
double root because the $x^2$ term in $\widetilde{y_m}^2$ vanishes and there
exists an overall factor $x^4$.
Therefore, we have $y_m^2 = \widetilde{y_m}^2$.
These subspaces 
are also on the unbroken $SO(4)$ branch corresponding to
$n=0$. At these points, the characteristic function 
$P_4(x) = x^2 \left( x^2 + 2 \eta \La^2\right)$
 is equal to $2 \rho^2 x^2 \La^2 {\cal T}_2 
\left(\frac{x}{2 \rho \La} \right)$ with $\rho^4=1$, by identifying $\eta = -
\rho^2$ (In other words, these are the vacua that survive when the 
${\cal N}=2$ theory is perturbed by a quadratic superpotential ($n=0$) and 
the $SO(4)$ gauge theory becomes massive at low energies) 
\footnote{We explicitly write some of the first 
Chebyshev polynomials ${\cal T}_{l}(x) $ where $l=1,2, \cdots, 6$ 
as follows:
\bea
{\cal T}_1(x) & = & x, \qquad
{\cal T}_2(x)  =  2x^2-1, \nonu \\
 {\cal T}_3(x) & = & 4x^3 - 3x, \qquad
 {\cal T}_4(x)  =  8x^4-8x^2 +1, \nonu \\
 {\cal T}_5(x) & = & 16x^5 - 20x^3+
5x, \qquad
{\cal T}_6(x)  =  32x^6-48x^4 +18 x^2 -1.
\nonu
\eea}.
Therefore,  this is the agreement with 
the glueball approach here exactly.

%%%%%%%%%%%%%%%%%%%%
$\bullet$ $SO(5)$
%%%%%%%%%%%%%%%%%%%%

In $SO(5)$ case, there is a breaking pattern 
$(N_0,N_1)=(3,1)$. 
In this case, 
one predicts 
the matrix model curve, by putting $(N,N_1)=(5,1)$ 
(note that $N=N_0+2N_1$), 
\bea
y_m^2 = x^2 \left( x^2+ m \right)^2 - 4   \La^6=
\left( x^2 + \eta \La^2  \right)^2 \left(x^2 + 4 \eta \La^2 \right)
\nonu
\eea
with $m = 3 \eta \La^2$ where $\eta^3=-1$.
For $SO(2M+1)$ case, we take a minus sign 
inside the log (\ref{solution2}).
That is, $\eta^{N-2}=-1$ for $N$ odd. 
We  find $\langle S \rangle= 0$. In \cite{ao}, 
the factorization problem 
resulted in the matrix model curve 
$
\widetilde{y_m}^2 = x^2 \left( x^2- l^2 \right)^2 - 4  \La^6 
$ and one can easily check that the intersections with the $n=0$ branch 
occur at $l^2 = -3 \eta \La^2$ where the additional double root appears
in $\widetilde{y_m}^2$. At these points, the characteristic function 
$P_4(x)=x^2 \left( x^2 + 3 \eta \La^2 \right)$ can be written as
$2 \rho^3 x \La^3 {\cal T}_3 
\left(\frac{x}{2 \rho \La} \right)$ with 
$\rho^6=1$, by identifying $\eta = -
\rho^2$. 
Therefore,  this is  the agreement with the glueball approach here.
The $\widetilde{f_0}$ is  equal to 
$\langle f_0 \rangle$.

%%%%%%%%%%%%%%%%%%%
$\bullet$ $SO(6)$
%%%%%%%%%%%%%%%%%%

For $SO(6)$ case, we consider two 
breaking patterns characterized by $(N_0,N_1)=(2,2)$ and $(4,1)$. 
In this case from 
the relations (\ref{sol}), (\ref{sol1}) and (\ref{formula}), 
one predicts 
the matrix model curve, by putting $(N_0,N_1)=(2,2)$, 
\bea
y_m^2 = x^2 \left( x^2+ m \right)^2 - 4  \eta^2 \La^4 x^2=
x^4 \left( x^2 + 4 \eta \La^2 \right)
\nonu
\eea
with $m = 2 \eta \La^2$ 
and $ \langle S \rangle=\eta^2 \La^4$, where $\eta$ is 4-th root of 
unity. 
From the results 
in \cite{ao}, the value of glueball is given 
as $\widetilde{S}=-\epsilon^2 \Lambda^4$, where $\ep$ is 4-th root of unity. 
This is the agreement with the glueball approach here
by identifying  $\ep^2 =-\eta^2$ precisely.
The factorization problem \cite{ao}
resulted in the matrix model curve 
$
\widetilde{y_m}^2 = x^2 \left[ \left( x^2- a^2+ \ep \sqrt{2} \La^2 \right)^2 +
4 \ep^2 \La^4 \right]
$. There were some typos in \cite{ao}.
The intersections with the $n=0$ branch 
occur at $a^2 =  \ep \sqrt{2} \La^2- 2 \eta \La^2$ 
where $\widetilde{y_m}^2$ has two double roots at $x=0$.
At these points, 
the characteristic function 
$P_6(x) = x^2 \left( x^2 -  \ep \sqrt{2} \La^2 + 2 \eta \La^2 \right)
\left( x^2 +  \ep \sqrt{2} \La^2 + 2 \eta \La^2   \right)$
is equal to $2 \ep^2 x^2 \La^4 {\cal T}_2 
\left(\frac{P_4(x)}{2 \ep x^2 \La^2 } \right)$ with $\ep^4=1$
where $P_4(x)=x^2\left( x^2+ 2\eta \La^2 \right)$.
This branch was
constructed by multiplication map by $K=2$
of $P_4(x)$. 
How do we know this is the solution of $n=0$ branch?
One can write down 
$P_6(x) - 2 \eta \La^4 x^2  =x^2 \left( x^2 + 2\eta \La^2 \right)^2
$ and
$P_6(x) + 2 \eta \La^4 x^2  =x^2 \left( x^2 + 2\eta \La^2 \right)^2
-4 \eta \La^4 x^2
$. Then the first branch has a single double root
and the second branch can be written as 
$ x^4 \left( x^2 + 4 \eta \La^2 \right)
$ which has an extra double root.
Therefore, these points are on the branch with $n=0$
and unbroken $SO(6)$.
These are the vacua that survive when the 
${\cal N}=2$ theory is perturbed by a quadratic superpotential ($n=0$) and 
the $SO(6)$ gauge theory becomes massive at low energies.
 
On the other hand, for other breaking pattern,
one predicts 
the matrix model curve, by putting $(N_0,N_1)=(4,1)$, 
\bea
y_m^2 = x^2 \left( x^2+ m \right)^2 + 4  \eta^2 \La^4 x^2 +
16 \eta^3 \La^6 =
\left( x^2 + 2\eta \La^2  \right)^2 \left( x^2 + 4\eta \La^2  \right)
\nonu
\eea
with $m = 4 \eta \La^2$ 
and $ \langle S \rangle =-\eta^2 \La^4$, 
where $\eta$ is 4-th root of 
unity. 
From the results 
in \cite{ao}, the value of glueball is given 
as $\widetilde{S}=-\epsilon \Lambda^4$, where $\ep$ is 2-nd root of unity. 
This is  the agreement with the glueball approach here
by identifying with $\ep =\eta^2$.
The factorization problem \cite{ao}
resulted in the matrix model curve 
$
\widetilde{y_m}^2 = x^2  \left( x^2 +A \right)^2 +
4 \ep \La^4 \left( x^2 + A \right) 
$. 
The intersections with the $n=0$ branch 
occur at $A =  4 \eta \La^2$ 
where $\widetilde{y_m}^2$ has an additional 
double root because $\widetilde{y_m}^2$ contains $\left( x^2 + 2 \eta 
\La^2 \right)^2$.
At these points, 
the characteristic function 
$P_6(x) = x^2 \left( x^4 +   4 \eta \La^2 x^2+  2 \eta^2 \La^4\right)$
 is equal to $2 \rho^4 x^2 \La^4 {\cal T}_4 
\left(\frac{x}{2 \rho \La} \right)$ with $\rho^8=1$, 
by identifying $\eta = -
\rho^2$. These are the vacua that survive when the 
${\cal N}=2$ theory is perturbed by a quadratic superpotential ($n=0$) and 
the $SO(6)$ gauge theory becomes massive at low energies.

%%%%%%%%%%%%%%%%%%%
$\bullet$ $SO(7)$
%%%%%%%%%%%%%%%%%%%

As in \cite{ao}, we consider two breaking 
patterns described by $(N_0,N_1)=(3,2)$ and $(5,1)$. 
For $(N_0,N_1)=(3,2)$,
the matrix model curve, by putting $(N_0,N_1)=(3,2)$,
implies 
\bea
y_m^2  & = & x^2 \left( x^2 + m \right)^2 +2 \eta^2 \left(1-
\sqrt{5} \right) \La^4 x^2 +2 \eta^3 \left( 7-3\sqrt{5} \right) \La^6
\nonu \\
& = & \left( x^2 + 4\eta \La^2  \right) 
\left[ x^2 + \left( \frac{3-\sqrt{5}}{2} \right) \eta \La^2 \right]^2
\nonu
\eea
with 
$m=\frac{\left(7-\sqrt{5} \right) \eta \La^2}{2}$ and 
$\eta^5=-1$.
We also find $\langle S \rangle=-\frac{1}{2} \eta^2 \left(1-
\sqrt{5} \right) \La^4 $. In \cite{ao}, the factorization problem 
turned out the matrix model curve 
$ \widetilde{y_m}^2=x^2 \left( x^2 -A +\frac{\ep^2 \La^5}
{A^{3/2}}\right)^2 -\frac{4\ep^2 \La^5}{A^{1/2}} x^2 -
\frac{4\La^{10}}{A^2}
$ with $\ep^4=1$. There were some typos in \cite{ao}.
The intersections with the $n=0$ branch occur at 
$A^2=\frac{7+3\sqrt{5}}{2} \La^4 \eta^2$ where the additional extra
double root appears in $\widetilde{y_m}^2$ with $\ep^{4/5}=-\eta$.
At these points, the characteristic function $P_{6}(x)=
x^2 \left( x^2 -a  \right)\left( x^2 -a -b\right)
$ given in \cite{ao}
is  written as  $2 \rho^5 x \La^5 {\cal T}_5 
\left(\frac{x}{2 \rho \La} \right)$ with 
$\rho^{10}=1$, by identifying $\eta = -
\rho^2$. 
These are the vacua that survive when the 
${\cal N}=2$ theory is perturbed by a quadratic superpotential ($n=0$).

For $(N_0,N_1)=(5,1)$,
the matrix model curve, by putting $(N_0,N_1)=(5,1)$,
implies 
\bea
y_m^2 & = & x^2 \left( x^2 + m \right)^2 +2 \eta^2 \left(1+
\sqrt{5} \right) \La^4 x^2 +2 \eta^3 \left( 7+3\sqrt{5} \right) \La^6
\nonu \\
& = & \left( x^2 + 4\eta \La^2  \right) 
\left[ x^2 + \left( \frac{3+\sqrt{5}}{2} \right) \eta \La^2 \right]^2
\nonu
\eea
with 
$m=\frac{\left(7+\sqrt{5} \right) \eta \La^2}{2}$.
The intersections with the $n=0$ branch occur at 
$A^2=\frac{7-3\sqrt{5}}{2} \La^4 \eta^2$ where the additional extra
double root appears in $\widetilde{y_m}^2$.
At these points, the characteristic function 
$P_{6}(x)$
is equal to $2 \rho^5 x \La^5 {\cal T}_5 
\left(\frac{x}{2 \rho \La} \right)$ with 
$\rho^{10}=1$, by identifying $\eta = -
\rho^2$.

%%%%%%%%%%%%%%%%%%
$\bullet$ $SO(8)$
%%%%%%%%%%%%%%%%%

Next we move to $SO(8)$ case. As in \cite{ao}, we consider two breaking 
patterns described by $(N_0,N_1)=(4,2)$ and $(6,1)$. 
For $(N_0,N_1)=(4,2)$,
the matrix model curve, by putting $(N_0,N_1)=(2,2)$,
implies 
\bea
y_m^2 = x^2 \left( x^2+ m \right)^2 + 4  \eta^3 \La^6 =
\left( x^2 + \eta \La^2  \right)^2 \left( x^2 + 4 \eta \La^2 \right)
\nonu
\eea
with $m = 3 \eta \La^2$ 
and $ \langle S \rangle=0$, where $\eta$ is 6-th root of 
unity. 
On the confining branch, the
value of $\widetilde{S}$ is zero and $\widetilde{f_0} = 4 
\ep \La^6$ where $\ep =\pm 1$. 
This is the agreement with the glueball approach here
by putting $\ep =\eta^3$. 
The factorization problem \cite{ao}
resulted in the matrix model curve 
$
\widetilde{y_m}^2 = x^2  \left( x^2 -a^2 \right)^2 +
4 \ep \La^6 
$. 
The intersections with the $n=0$ branch 
occur at $a^2 =  -3 \eta  \La^2$ 
where $\widetilde{y_m}^2$ has an additional 
double root because $\widetilde{y_m}^2$ contains $\left( x^2 + \ep^{1/3} 
\La^2 \right)^2$.
At these points, 
the characteristic function 
$P_8(x) = x^4 \left( x^2 + 3 \eta  \La^2 \right)^2  + 2 \eta^3 \La^6 x^2 
$
can be written as  $2 \rho^6 x^2 \La^6 {\cal T}_6 
\left(\frac{x}{2 \rho \La} \right)$ with $\rho^{12}=1$, 
by identifying $\eta = -
\rho^2$. These are the vacua that survive when the 
${\cal N}=2$ theory is perturbed by a quadratic superpotential ($n=0$) and 
the $SO(8)$ gauge theory becomes massive at low energies.

On the other hand, on the Coulomb 
branch $\widetilde{S}$ is parametrized by $a$. 
Let us consider the breaking pattern $SO(8) \to SO(6) \times U(1)$
where we have $(N,N_1)=(8,1)$.
One can write down the matrix model curve
as
\bea
y_m^2 = x^2 \left( x^2 +m \right)^2 + 8 \eta^2 \La^4 x^2 + 36 \eta^3 \La^6=
\left(x^2 + 3\eta \La^2 \right)^2 \left( x^2 + 4 \eta \La^2 \right)
\nonu
\eea
with $m=5 \eta \La^2$ where $\eta^6=1$ and $\langle S \rangle =-2\eta^2 
\La^4 $.
The factorization problem \cite{ao} turned out
the matrix model curve 
 $
\widetilde{y_m}^2 = x^2  \left( x^2 +\frac{4 \ep \La^6}{a^4} -
a^2 \right)^2 -\frac{8\ep \La^6}{a^2} x^2 + \frac{4\ep \La^6}{a^2}
\left( a^2-\frac{8\ep \La^6}{a^4} \right)
$ with $\ep^2=1$ and the glueball $\widetilde{S}$ becomes 
$\frac{2\ep \La^6}{a^2}$ implying $a^2=-\frac{\ep}{\eta^2} \La^2$.
Since we are interested in the special 
point $n=0$ and $n=1$ singularity we 
should constrain one more double root for the 
matrix model curve. 
The intersections with the $n=0$ branch
occur at $\frac{4 \ep \La^6}{a^4} -
a^2= 5\eta \La^2$. Therefore, it leads to the value $a^2 = -\eta \La^2$
by identifying $\ep =\eta^3$. At these points, the characteristic function
$P_8(x)=x^2\left(x^2-a^2 \right)^2 \left( x^2 + \frac{4\ep \La^6}{a^4} \right)-
2\ep \La^6 x^2=x^2 \left(x^2 + \eta \La^2 \right)^2 \left( x^2 +
4 \eta \La^2 \right)-2\eta^3 \La^6 x^2$ can be written as
 $2 \rho^6 x^2 \La^6 {\cal T}_6 
\left(\frac{x}{2 \rho \La} \right)$ with $\rho^{12}=1$, 
by identifying $\eta = -
\rho^2$. 

%It give $a^6=-\eta \La^6$, where $\eta$ is 2-nd root of unity. 
%Therefore if we represent 3-rd root of unity $(\eta)^{\frac{2}{3}}$ 
%as $\epsilon$ we find $S=-2\epsilon \La^4$. We can see agreement 
%if we put $(N_0,N_1)=(6,1)$ into $(\ref{formula})$. 

It was noticed in \cite{ao} that 
there exists also a Coulomb branch where the $SO(8)$ breaks into
$SO(4) \times U(2)$. That is, $(N_0,N_1)=(4,2)$.
We want to show that for given matrix model curve, 
$(N_0,N_1)$ are determined uniquely as follows 
\footnote{We are grateful to
D. Shih \cite{Shih} for relevant discussion on the $U(4)$ gauge theory in the 
Coulomb branch where although there exist two classical limits, $U(4) \to
U(2) \times U(2)$ and $U(4) \to U(3)\times U(1)$, 
only the latter can be applied to the matrix model curve 
at $n=1$ and $n=2$ singularity through the computations on 
$(N_1,N_2)$.  }.
In order to see the precise values $(N_0,N_1)$ on the Coulomb branch 
we calculate them by putting $\eta=-1,\La=1$, and $a=1$, that are consistent 
with the previous paragraph, 
into the expressions given in \cite{ao}. 
The results can be rewritten as 
\begin{eqnarray}
\widetilde{y_m}^2=
\left(x^2-4 \right) \left(x^2-3 \right)^2,
\qquad  P_8(x)=
x^2 \left(x^2 -1 \right)^2 \left( x^2 -
4 \right)+2  x^2. 
\label{6sep1}
\end{eqnarray}
By using these relations, the function $T(x)=\mbox{Tr} \frac{1}{x-\Phi}$ 
is given as \cite{ao} 
\begin{eqnarray}
T(x)=\frac{P_8^{\prime}(x)-\frac{2}{x}P_8(x)}{\sqrt{P_8(x)^2-
4x^4\La^{12}}}+\frac{2}{x}=\frac{6}{\sqrt{x^2-4}}+\frac{2}{x}.
\nonu
\end{eqnarray}
As we can see from the first equation of (\ref{6sep1}), there 
exist three branch cuts on the $x$-plane $[-2, -\sqrt{3}],
[-\sqrt{3}, \sqrt{3}]$, and $[\sqrt{3}, 2]$. Since we are assuming 
$n=0$ and $n=1$ singular case, these three 
branch cuts before taking the limit
are joined at the 
locations of $x= \pm 
\sqrt{3}$ after taking the limit and they become a single 
branch cut $[-2,2]$. 
Therefore, we can explicitly calculate $(N_0,N_1)$ as 
follows through (\ref{n0n1}):
\begin{eqnarray}
N_0&=& \frac{1}{2\pi i} \oint_{A_0} T(x) dx=
 \frac{2}{2\pi i}\int_{-\sqrt{3}}^{\sqrt{3}}\left
(\frac{6}{\sqrt{x^2-4}}+\frac{2}{x} \right)dx=\left(\frac{12}
{\pi}\int_0^{\sqrt{3}}\frac{dx}{\sqrt{4-x^2}}\right)+2 =6, \nonu 
\\
%& = & \frac{12}{\pi}
%\sin^{-1}\left( \frac{\sqrt{3}}{2}\right) +2=6, \nonu \\
N_1&=&
 \frac{1}{2\pi i} \oint_{A_1} T(x) dx=
\frac{2}{2\pi i}\int_{\sqrt{3}}^{2}\left(\frac{6}
{\sqrt{x^2-4}}+\frac{2}{x} \right)dx=\frac{6}{\pi} 
\int_{\sqrt{3}}^{2}\frac{dx}{\sqrt{4-x^2}}=1.
\nonu
\end{eqnarray}
where we used the theorem of residue around the origin:
$\frac{1}{2\pi i} \oint_{A_0} \frac{1}{x} dx=1$. 

Although there exist two classical limits, $SO(8) \to SO(4) \times U(2)$
and $SO(8) \to SO(6) \times U(1)$, at the $n=0$ and $n=1$ 
singularity, one must use one of them, $SO(8) \to SO(6) \times U(1)$.
Let us emphasize that $N_0$ and $N_1$ are defined to be the on-shell 
periods of the one-form $T(x)$ at the $n=0$ and $n=1$ singularity.

%What is the effective glueball superpotential in terms of 
%$f_0$ and $f_2$?(!!!!!!!!!!!!!!!!!!!!!!!!!!!!!!)

%%%%%%%%%%%%%%%%%%%%%%%%%%%%%%%%%%%%%%%%%%%%%%%%%%%%%%%%%%%%%%%%%%%%%%%%%%%%%
\subsection{The coupling constant near the singularity}
%\setcounter{equation}{0}
%%%%%%%%%%%%%%%%%%%%%%%%%%%%%%%%%%%%%%%%%%%%%%%%%%%%%%%%%%%%%%%%%%%%%%%%%%%%%

\indent

To see the matrix of coupling constant near the singularity we 
consider the effective superpotential from which we can read off
 the matrix of coupling constant for $U(1)^n$ gauge groups. 
As already discussed in \cite{cdsw,oz,ashoketal,jo,ow}, 
the effective superpotential 
is given as, by considering $RP^2$ contribution,
\begin{eqnarray}
W_{\mbox{eff}}=\int d^2 \psi {\cal F}_p +4 {\cal F}_{RP^2},
\qquad
{\cal F}_{RP^2}=- \frac{1}{2} \frac{\partial {\cal F}_{p} }
{\partial S_0}
\nonu
\end{eqnarray}
for some function ${\cal F}_p$.
By performing the $\psi$ integrals,
expanding 
in powers of $w_{\al i}$, and 
collecting  the terms proportional to $w_{\alpha}w^{\alpha}$, one obtains
\begin{eqnarray}
W_{\mbox{eff}}\sim \frac{1}{2}\sum_{ij}\frac{\partial^2 {\cal F}_p}
{\partial S_i \partial S_j}w_{\alpha i}w^{\alpha}_j -\frac{N_0 - 2}{2N_l} 
\frac{\partial^2 {\cal F}_p}{\partial S_0 
\partial S_l}w_{\alpha l}w^{\alpha}_l-\sum_{i,k} \frac{N_i}{2N_k} 
\frac{\partial^2 {\cal F}_p}{\partial 
S_i \partial S_k}w_{\alpha k}w^{\alpha}_k
\nonu
\end{eqnarray}
where $i, j, k=1,2,\cdots, n$. Since the $SO(N_0)$ group 
does not have an $U(1)$
factor, the corresponding $U(1)$ gauge field does not exist.
The $w_{\al i}$'s come from the $U(1)$ factor in $U(N_i) =U(1) 
\times SU(N_i)$.
The matrix of gauge couplings is given by
the formula \cite{cdsw,ow,oz}
\begin{eqnarray}
\frac{1}{2\pi i}\tau_{ij}=\frac{\partial^2 
{\cal F}_p(S_k)}{\partial S_i \partial S_j}-\delta_{ij}
\frac{1}{N_i}\sum_{l=1}^n N_l \frac{\partial^2 {\cal F}_p(S_k)}
{\partial S_i \partial S_l}-\delta_{ij} \left(\frac{N_0 - 2}{N_i}
\right)
\frac{\partial^2 {\cal F}_p(S_k)}{\partial S_0\partial S_i},
\;\; i, j=1,2, \cdots, n.
\nonu
\end{eqnarray}
In the quartic tree-level superpotential case ($n=1$), 
there is only one coupling constant and it is
given by, due to the cancellation of first two
terms above, 
\begin{eqnarray}
\frac{1}{2\pi i}\tau=-\left( \frac{N_0- 2}{N_1} \right)
\frac{\partial \Pi_1}
{\partial S_0}=
- \frac{i \pi}{16}  \frac{\left(N_0-2 \right)^2}{ \left(N-2 \right)^2} 
\frac{1}{ \log \left(\frac{16}{1-k^{\prime 2}}\right)}.
\nonu
\end{eqnarray}
Here $k^{\prime}$ is defined as (\ref{def}).
Recall that the $\Pi_1$ is a derivative of the prepotential 
with respect to $S_1$: $\frac{\pa {\cal F}_p(S_k)}{\pa S_1 }$.
The single derivative $\frac{\partial \Pi_1}
{\partial S_0}$ was evaluated at the extremum of the 
effective superpotential.
We used the result of Appendix A (\ref{coupling}). 
%where we use notation $SO(N) \to SO(N_0)\times U(N_1)$.
So $\tau$ goes to zero as one approaches the $n=0$ and $n=1$
singularity. The $\tau$ is continuous as we move from the $n=1$
branch to the $n=0$ branch since the $\tau$ is zero on the $n=0$
branch.
The logarithmic behavior implies that the gauge coupling constant 
of the nontrivial $U(1)$ diverges as we approach the singularity.
See also the similar behavior in \cite{fm}.
This divergence comes from the additional monopole that becomes massless 
at the singularity.

%%%%%%%%%%%%%%%%%%%%%%%%%%%%%%%%%%%%%%%%%%%%%%%%%%%%%%%%%%%%%%%%%%%%%%%%%%%%%
\section{The $n=0$ and $n=1$ singularity: $USp(2N)$ gauge theory}
\setcounter{equation}{0}
%%%%%%%%%%%%%%%%%%%%%%%%%%%%%%%%%%%%%%%%%%%%%%%%%%%%%%%%%%%%%%%%%%%%%%%%%%%%%

%%%%%%%%%%%%%%%%%%%%%%%%%%%%%%%%%%%%%%%%%%%%%%%%%%%%%%%%%%%%%%%%%%%%%%%%%%%%%
\subsection{The glueball equations of motion}
\setcounter{equation}{0}
%%%%%%%%%%%%%%%%%%%%%%%%%%%%%%%%%%%%%%%%%%%%%%%%%%%%%%%%%%%%%%%%%%%%%%%%%%%%%

\indent

Now we discuss the intersections of the $n=0$ and $n=1$
branches for $USp(2N)$ gauge theory. 
A vacuum with unbroken gauge group 
$USp(2N_0) \times U(N_1)$ meets a vacuum with unbroken 
gauge group $USp(2N)$ with
\bea
2N=2N_0 + 2N_1.
\nonu
\eea
These intersections occur at the particular values of the tree-level
superpotential parameter.
One can proceed 
the method given in previous section similarly.
The matrix model curve is given in (\ref{matrixcurve})
and the tree-level superpotential has a parameter $m$: $W_3^{\prime}(x)=
x^3 + m x$.
The periods and their conjugate periods are 
written as those in $SO(N)$ case
and they provide the effective superpotential \cite{feng} 
\begin{eqnarray}
W_{\mbox{eff}}=  2 \pi i \left[ \left(N_0 + 2 \right) \Pi_0 +2N_1 \Pi_1
\right] -
2 \left(2N + 2 \right) S \log 
\left(\frac{\Lambda}{\Lambda_0} \right). 
\label{superpotentialusp}
\end{eqnarray}
There is no $b_1$ term.
Here $N_0$ and $N_1$ are defined by
(\ref{n0n1}) and the corresponding operator $T(x)$ for $USp(2N)$ gauge 
theory is \cite{ao}
\bea
T(x) & = & \frac{d}{d x} \log \left[ B_{2N+2}(x) + \sqrt{B_{2N+2}^2(x)- 4 
\La^{4N+4}} -\log x^2  \right], \nonu \\
B_{2N+2}(x) & = & x^2 P_{2N}(x) + 
2 \La^{2N+2}, \qquad P_{2N}(x) = \mbox{det} \left( x-\Phi \right).
\nonu
\eea
The derivatives of $\Pi_i$ with respect to $f_j$ are the same as 
those in $SO(N)$ case exactly 
and by using the equations of motion for $f_0$
and $f_2$ together with the effective superpotential 
(\ref{superpotentialusp}),
one obtains
two relations
\begin{eqnarray}
(N+1)\log \Bigg| \frac{x_2^2}{4\Lambda^2} \Bigg|=0,\qquad
-\left(\frac{x_2^2-2x_0^2}{x_2^2} \right)=\cos 
\left(\frac{2\pi N_1}{2N+2} \right)
\nonu
\end{eqnarray}
where $\eta$ is $(N+1)$-th root of unity.
%(This can be related to 
%the Veneziano-Yankielowicz term?!!!!!!!!!!!!!). 

From these two equations we obtain the following results
with (\ref{constraint}),
\begin{eqnarray}
x_2^2=4\eta \Lambda^2,
\qquad x_0^2=2\eta \La^2 \left(1+\cos 
\frac{2\pi N_1}{2N+2} \right), \qquad
m= 2\eta \Lambda^2 
\left( 2+\cos \frac{2\pi N_1}{2N+2} \right)
\nonu
\end{eqnarray}
where $\eta^{N+1}=1$.
Note that compared with the $SO(N)$ gauge theory, 
the $N$ dependence appears in the denominator of cosine function
differently and the property of the phase factor $\eta$.  
By using the relation between the two parametrizations,
one obtains the expectation values of the fields $f_2, f_0$ and $S$
at the double root singularity
\begin{eqnarray}
\langle f_2 \rangle &=& 4\eta^2 \La^4 \left( 1+2\cos 
\frac{2\pi N_1}{2N+2} \right), \nonu \\
\langle f_0 \rangle &=& 16\eta^3 \La^6 \left( 1+\cos 
\frac{2\pi N_1}{2N+2} \right)^2,  \nonu \\
\langle S \rangle &=& -\eta^2 \La^4 \left( 1+2\cos 
\frac{2\pi N_1}{2N+2} \right).
\nonu
\end{eqnarray}
Also we write the matrix model curve as
\bea
y_m^2 & = & x^2 \left[ x^2+      
 2 \eta \La^2  \left( 2+ c \right) \right]^2 +
 4 \eta^2 \Lambda^4 \left(1+2 c 
\right) x^2 
 + 
16 \eta^3 \La^6 \left( 1+ c \right)^2 
\nonu \\  
&= & \left[x^2 + 2\eta \Lambda^2 
\left( 1+ c \right) \right]^2 
\left( x^2 + 4\eta \Lambda^2  \right), \qquad c \equiv 
\cos \left( \frac{2\pi N_1}{2N+2} \right). 
\label{for}
\eea

To demonstrate this general results
one can compare the formula (\ref{for}) 
with the explicit examples given \cite{ao}
by imposing the additional condition for an extra double root. 
Let us consider $USp(2), USp(4)$ and $USp(6)$.

%%%%%%%%%%%%%%%%%%%%%%
$\bullet$ $USp(2)$
%%%%%%%%%%%%%%%%%%%%%

One predicts the matrix model curve by inserting 
$(N,N_1)=(1,1)$
\bea
y_m^2 =x^2 \left( x^2 + m \right)^2 +  4 \La^4 x^2 +
 16\eta \La^6 = \left( x^2 + 2\eta \La^2 \right)^2 \left( x^2 + 
4 \eta \La^2 \right) 
\nonu
\eea 
with $m=4 \eta \La^2$ and 
\begin{eqnarray}
\langle f_2 \rangle = 4 \La^4 ,\qquad  \langle f_0 \rangle = 
16\eta \La^6 ,\qquad \langle S \rangle = -\La^4
\nonu 
\end{eqnarray}
where $\eta$ is $2$-nd root of unity. 
The factorization problem \cite{ao} resulted in 
the matrix model curve $
\widetilde{y_m}^2 = x^2 \left( x^2- v^2 \right)^2 + 4 \La^4 \left( 
x^2 -v^2\right)
$. There were some typos in \cite{ao}.
The intersections with the $n=0$ branch occur at $v^2 =-4\eta \La^2$
where 
$\widetilde{y_m}^2$ has an additional 
double root. 
At these points, the characteristic function $B_4(x)=x^2\left(x^2+
4 \eta \La^2  \right)+2 \La^4$ is equal to $2 \rho^2 \La^4 {\cal T}_2
\left( \frac{x^2}{2 \rho \La^2} +1\right)$ with $\rho^2 =1$ 
by identifying $\rho=\eta$. Recall that the function $B_4(x) \equiv
x^2 P_2(x) + 2 \La^{4}$. 
These are the vacua that survive when the ${\cal N}=2$ theory is
perturbed by a quadratic superpotential ($n=0$).
These subspaces are on the unbroken $USp(2)$ branch.
This is an agreement with the glueball approach.

%%%%%%%%%%%%%%%%%%%%
$\bullet$ $USp(4)$
%%%%%%%%%%%%%%%%%%%%

At first, we consider the breaking pattern $USp(4) \to U(2)$, 
namely $N=2,N_1=2$ and $\eta^3=1$. Putting these results, we obtain 
\begin{eqnarray}
\langle f_2 \rangle = 0 ,\qquad  \langle f_0 \rangle = 4 \La^6 ,\qquad 
\langle S \rangle = 0, 
\nonu
\end{eqnarray}
where the matrix model curve 
becomes
\bea
y_m^2 = x^2 \left(x^2 +m \right)^2 + 4 \La^6=
\left( x^2 + \eta \La^2 \right)^2 \left( x^2 + 4\eta \La^2 \right)
\nonu
\eea
with $m=3 \eta \La^2$.
From the results of  \cite{ao}, the glueball field 
$\widetilde{S}$  vanishes. The factorization problem turned out
$\widetilde{y_m}^2=x^2\left(x^2 -a^2 \right)^2 +4\La^6$. 
Without any difficulty the intersections with the $n=0$ branch
occur at $a^2=-3\eta \La^2$ where the additional double root 
appears in $\widetilde{y_m}^2$.
In this case, 
the characteristic function $B_6(x)=x^2 \left( x^2 +3 \eta \La^2 \right)^2 
+2\La^6$ can be written as
$2 \rho^3 \La^6 {\cal T}_3
\left( \frac{x^2}{2 \rho \La^2} +1\right)$ with $\rho^3 =1$ 
by identifying $\rho=\eta$.  
These are the vacua that survive when the ${\cal N}=2$ theory is
perturbed by a quadratic superpotential and the $USp(4)$ gauge theory 
becomes massive at low energies.

Next breaking 
pattern is $USp(4)\to USp(2)\times U(1)$. That is, $N=2,N_1=1$,
one obtains 
\begin{eqnarray}
\langle f_2 \rangle = 8\eta^2 \La^4 ,\qquad  \langle f_0 \rangle = 
36\La^6 ,\qquad \langle S \rangle = -2\eta^2\La^4
\nonu 
\end{eqnarray}
and the matrix model curve
is
\bea
y_m^2 = x^2 \left(x^2 +m  \right)^2 + 8\eta^2 \La^4 x^2 + 36\La^6 
= \left( x^2 + 3 \eta \La^2  \right)^2 \left( x^2 + 4\eta \La^2  \right)
\nonu
\eea
with $m=5 \eta \La^2$.
The matrix model curve $\widetilde{y_m}^2=\left(x^2 + 
\frac{4\La^6}{a^4} \right) \left[ \left(x^2 -a^2 \right)^2 +\frac{4\La^6}{a^4} 
\left( x^2 -2 a^2 \right) \right]  $ 
written in \cite{ao} has an extra double root 
when $a^2=-\eta \La^2$ and $\widetilde{S}=\frac{2\La^6}{a^2}$.
The intersections with the $n=0$ branch occur at $a^2=-\eta \La^2$ and
the characteristic function $B_6(x)=\left( x^2 +\eta \La^2 \right)^2 
\left(  x^2 + 4 \eta \La^2 \right)-2\La^6$ becomes
$2 \rho^3 \La^6 {\cal T}_3
\left( \frac{x^2}{2 \rho \La^2} +1\right)$ with $\rho^3 =1$ 
by identifying $\rho=\eta$.  
%Putting these values in the result we find that $f(x)=8\epsilon^3 
%\La^4x^2+36\La^6$. Identifying $\epsilon$ as $\eta$ in the result by 
%glueball approach we found exact agreement.

%What about $(s_{+}, s_{-})=(1,0)$(!!!!!!!!!!)
In order to see the precise values $(N_0,N_1)$ on the Coulomb branch 
we calculate them by putting $\eta=1, \La=1$, and $a=1$ 
that are also consistent with the previous paragraph, 
into the expressions given in \cite{ao}. 
The results can be rewritten as 
\begin{eqnarray}
\widetilde{y_m}^2=
\left(x^2+4 \right) \left(x^2+3 \right)^2,
\qquad  B_6(x)=
\left(x^2+1 \right)^2 \left(x^2+4 \right)-2. 
\label{6sep2}
\end{eqnarray}
By using these relations, the function $T(x)$ is given as \cite{ao}
\begin{eqnarray}
T(x)=\frac{B_6^{\prime}(x)}{\sqrt{B_6(x)^2-
4\La^{12}}}-\frac{2}{x}=\frac{6}
{\sqrt{x^2+4}}-\frac{2}{x}.
\nonu
\end{eqnarray}
As we can see from the first equation of (\ref{6sep2}), there 
are three branch cuts on the $x$ plane $[ -2 i,-\sqrt{3} i],
[ -\sqrt{3} i, \sqrt{3} i]$, and $[ \sqrt{3} i, 2i]$. 
Since we are assuming 
$n=0$ and $n=1$ singular case, these branch cuts are joined at $\pm 
\sqrt{3} i$. Therefore, we can explicitly calculate $(N_0,N_1)$ as 
follows:
\begin{eqnarray}
2N_0&=&\frac{2}{2\pi i}\int_{-\sqrt{3} i}^{\sqrt{3} i}\left
(\frac{6}{\sqrt{x^2+4}}-\frac{2}{x} \right)dx=\left(\frac{12}
{\pi}\int_0^{\sqrt{3}}\frac{dx}{\sqrt{4-x^2}}\right)-2
%=\frac{12}{\pi}
%\sin^{-1}\left( \frac{\sqrt{3}}{2}\right) -2
=2, \nonu \\
N_1&=&\frac{2}{2\pi i}\int_{\sqrt{3} i}^{2 i}
\left(  \frac{6}
{\sqrt{x^2+4}}
-\frac{2}{x} \right)dx=\frac{6}{\pi} 
\int_{\sqrt{3}}^{2}\frac{dx}{\sqrt{4-x^2}}=1.
\nonu
\end{eqnarray}
where we used the residue theorem. 

Although there exist two different classical limits corresponding to
unbroken gauge group $USp(4) \to U(2)$
and $USp(4) \to USp(2) \times U(1)$, at the $n=0$ and $n=1$ 
singularity, one must use $USp(4) \to USp(2) \times U(1)$ since these
are the values of $(N_0,N_1)$ at the $n=0$ and $n=1$ singularities 
of these branches.

%%%%%%%%%%%%%%%%%%%%%
$\bullet$ $USp(6)$
%%%%%%%%%%%%%%%%%%%%

Let us consider the confining branch in which the gauge group breaks into 
$USp(6)\to USp(2)\times U(2)$.
Then one predicts the matrix model curve for $(N,N_1)=(3, 2)$ 
\bea
y_m^2 = x^2 \left( x^2 + m \right)^2 + 4\eta^2 \La^4  x^2 + 
16\eta^3 \La^6 =
\left( x^2 + 2\eta \La^2 \right)^2 \left( x^2 + 4\eta \La^2 \right)
\nonu
\eea
with $m=4\eta \La^2$ and  
\begin{eqnarray}
\langle f_2 \rangle = 4\eta^2 \La^4 ,\qquad  \langle f_0 \rangle = 
16\eta^3 \La^6 ,\qquad \langle S \rangle = -\eta^2\La^4,
\nonu 
\end{eqnarray}
where $\eta$ is $4$-th root of unity. 
The factorization problem turned out 
$\widetilde{y_m}^2=x^2 \left( x^2 -a -\frac{2 \ep \La^4}{a}\right)^2+
4 \ep \La^4 x^2 - 4 a \ep \La^4 -\frac{8\La^8}{a}$ with $\ep^2=1$ 
and the glueball field 
$\widetilde{S}=-\ep \La^4$ which is identical to $\langle S \rangle$ for 
$\ep =\eta^2$.
The matrix model 
curve written in \cite{ao} has an extra double root when 
$a=(-2 \pm \sqrt{2})\eta \La^2$. 
At these points, the characteristic function $B_8(x)=\left( x^2-
a \right)^2
\left( x^2 - \frac{2\ep \La^4}{a} \right)^2-2\La^8$ which is equal to
$2 \ep^2 \La^8 {\cal T}_2 \left( \frac{x^2 P_2(x)}{2 \ep \La^4} +1
\right)$
where $P_2(x) = x^2 - a -\frac{2\ep \La^4}{a}$.
Then how do we check these points are on the $n=0$ branch with unbroken 
$USp(6)$?
One can write down 
$B_8(x) - 2  \La^8  =x^2 \left( x^2 + 2\eta \La^2 \right)^2
\left(x^2 + 4 \eta \La^2  \right)
$ and
$B_8(x) + 2 \La^8  = \left( x^2\left( x^2 + 4\eta \La^2  \right)
 + 2 \La^4 \right)^2$. Then the first branch has an  extra double root
and the second branch has two  double roots.
Therefore, these points are on the branch with $n=0$
and unbroken $USp(6)$.

%Putting these values in the result 
%we find that $f_0=-16 (\sqrt{\eta})^3\La^6$. Since $\sqrt{\eta}$ is 
%$4$-th root of unity we can identify it as $\eta$ in the result by glueball 
%approach.

%%%%%%%%%%%%%%%%%%%%%%%%%%%%%%%%%%%%%%%%%%%%%%%%%%%%%%%%%%%%%%%%%%%%%%%%%%%%%
\subsection{The coupling constant near the singularity}
%\setcounter{equation}{0}
%%%%%%%%%%%%%%%%%%%%%%%%%%%%%%%%%%%%%%%%%%%%%%%%%%%%%%%%%%%%%%%%%%%%%%%%%%%%%

\indent
For $USp(2N)$ case, 
the matrix of gauge couplings is given by
the formula \cite{cdsw,ow,oz}, by taking the different contribution
to the ${\cal F}_{RP^2}$,
\begin{eqnarray}
\frac{1}{2\pi i}\tau_{ij}=\frac{\partial^2 
{\cal F}_p(S_k)}{\partial S_i \partial S_j}-\delta_{ij}
\frac{1}{N_i}\sum_{l=1}^n N_l \frac{\partial^2 {\cal F}_p(S_k)}
{\partial S_i \partial S_l}-\delta_{ij} \left(\frac{2N_0 + 2}{N_i}
\right)
\frac{\partial^2 {\cal F}_p(S_k)}{\partial S_0\partial S_i},
\;\; i, j=1,2, \cdots, n.
\nonu
\end{eqnarray}
The single gauge coupling for quartic superpotential is given similarly   
\begin{eqnarray}
\frac{1}{2\pi i}\tau=-\frac{\left(2N_0+ 2\right)}{N_1}\frac{\partial \Pi_1}
{\partial S_0}=
- \frac{i \pi}{16}  \frac{\left(2N_0+2 \right)^2}{ \left(2N+2 \right)^2} 
\frac{1}{ \log \left(\frac{16}{1-k^{\prime 2}}\right)}
\nonu
\end{eqnarray}
where we use a notation for $USp(2N) \to USp(2N_0)\times U(N_1)$.
As we have seen in previous consideration for $SO(N)$ case,
The $\tau$ is continuous as we move from the $n=1$
branch to the $n=0$ branch since the $\tau$ is zero on the $n=0$
branch.
The logarithmic behavior implies the gauge coupling constant 
of the nontrivial $U(1)$ diverges as we approach the singularity.

%%%%%%%%%%%%%%%%%%%%%%%%%%%%%%%%%%%%%%%%%%%%%%%%%%%%%%%%%%%%%%%%%%%%%%%%%%%%%
\section{The matrix model curve for degenerated case:$SO(N)$ gauge theory}
%\setcounter{equation}{0}
%%%%%%%%%%%%%%%%%%%%%%%%%%%%%%%%%%%%%%%%%%%%%%%%%%%%%%%%%%%%%%%%%%%%%%%%%%%%%

\indent

Contrary to non-degenerated case where every root of $W^{\prime}(x)$ 
has D5-branes wrapping around it,
in previous sections, there exists 
only one parameter denoted by $F$ for degenerated case. 
For the degenerated case \cite{afo1,afo2} 
where some roots of $W^{\prime}(x)$ do 
not have wrapping D5-branes around them, 
the matrix model curve is described as \cite{afo1,afo2} 
\begin{eqnarray}
y_{m,d}^2=\left(\frac{W_3^{\prime}(x)}{x} \right)^2+ 4F =
\left( x^2 + m \right)^2 + 4F \equiv 
\left(x^2+a^2 \right) \left(x^2+b^2 \right)
\nonu
\end{eqnarray}
where $W_3^{\prime}(x)=x^3 + m x$ as before. 
The first parametrization of this matrix model curve implies that
$m$ is a parameter and $4F$ is a fluctuating field 
that is related to the glueball field. The second parametrization 
in terms of the roots $\pm i a$ and $\pm i b$
will be convenient and is subject to the constraint
\bea
m=\frac{1}{2} \left( a^2 + b^2 \right).
\nonu
\eea

At first sight, one can derive the results for 
degenerate case by using the results in section
\ref{glueball equations}. 
The matrix model curve for the degenerated case 
can be represented by the matrix model curve $y_m$ (\ref{Curve}) corresponding
to  the non-degenerated case,
\begin{eqnarray}
y_{m}^2=x^2y_{m,d}^2=x^4 \left(x^2+4\eta \La^2 \right)
 \iff f_0=0.
\label{sep3general}
\end{eqnarray}
where $\eta^{N-2}=1$ for $N$ even  and $\eta^{N-2}=-1$ for $N$ odd
as before. 

As we will see below, this naive consideration gives 
the precise results %at the
%$n=0$ and $n=1$ singularity 
on the degenerated case. 
%Surprisingly we can obtain the
%general matrix model curve not only at $n=0$ and $n=1$ singularity 
%but also on the whole
%degenerated branch. 
On the degenerated case, although the matrix model curve
does not have singularity, the dynamical variables are less than
those in non-degenerated case. 
Therefore, the glueball approach for this case is very
powerful as the singular case and the equation of motions for the variables
becomes drastically easy to solve.    

The dual periods are given by the integrals for $x^2 y_{m,d}$ over $x$: 
\begin{eqnarray}
2\pi i \Pi_0&=& 
\int^{\Lambda_0}_{0}\sqrt{x^2 (x^2+a^2)(x^2+b^2)} \ dx, \nonumber \\
2\pi i \Pi_1&=&
\int^{i\Lambda_0}_{ib}\sqrt{x^2 (x^2+a^2)(x^2+b^2)} \ dx. 
\nonu
\eea
The corresponding effective glueball superpotential 
is given by (\ref{super}).
One can derive the matrix model curve by direct calculations from the 
effective superpotential for the degenerated case. Since there exists only 
one parameter, we have to consider 
the equation of motion of $F$ only.
After differentiating the dual periods with respect to the field $F$,
one obtains
\begin{eqnarray}
4\pi i \frac{\partial \Pi_0}{\partial F}&=&\int_0^{\La_0}
\frac{x^2dx}{\sqrt{x^2(x^2+a^2)(x^2+b^2)}}\simeq \frac{1}{2}
\log \Bigg| \frac{4\La_0^2}{(a+b)^2} \Bigg|,  \nonu \\
4\pi i \frac{\partial \Pi_1}{\partial F}&=&\int_{ib}^{i\La_0}
\frac{x^2dx}{\sqrt{x^2(x^2+a^2)(x^2+b^2)}}\simeq \frac{1}{2}
\log \Bigg|  \frac{4\La_0^2}{b^2-a^2} \Bigg|.
\nonu 
\end{eqnarray}
By using these two results and taking $\partial W_{\mbox{eff}}/
\partial F=0$, we obtain the following equation,
\begin{eqnarray}
\left[ \frac{4\La^2}{(a+b)^2}\right]^{N_0-2} \times 
\left( \frac{4\La^2}{b^2-a^2}\right)^{2N_1}= \pm 1. 
\label{5sepcondition}
\end{eqnarray}
It is noteworthy that  by taking the special limit, we can 
reproduce the equation
(\ref{sep3general}) because $b^2=4\eta \La^2, a=0$ and 
$y_{m,d}^2=x^2(x^2+4\eta \La^2)$.  However, in general, 
the relation (\ref{5sepcondition}) leads to the more general result. 

The matrix model curve 
for the degenerated case can be represented as 
\begin{eqnarray}
y_{m,d}^2
%=\left(x^2+\frac{a^2+b^2}{2} \right)^2-
%\frac{(b^2-a^2)^2}{4}\equiv 
=\left(x^2+m \right)^2+4F, \qquad
%\nonu
%\end{eqnarray}
%From the e.q.(\ref{5sepcondition}) and the condition 
%$(b^2-a^2)^2=16F$ we can rewrite $m$ as 
%
%\begin{eqnarray}
m
%=\frac{a^2+b^2}{2}
=\frac{K^2-16F}{4K},
\qquad 
K\equiv 
\left[  \frac{(-4\La^2)^{N-2}}{(-16F)^{N_1}} 
\right]^{\frac{1}{N_0-2}}
\label{formula2}
\end{eqnarray}
where $K$ becomes $\left( a+b \right)^2$ in the second parametrization.
This general formula provides the matrix model curve for degenerated case
and it depends on the parameter $F$, $N_0$ and $N_1$ where 
$N_0$ can be zero. Turning on the 
parameter $F$ to the special value will lead to the symmetry breaking 
$SO(N) \to SO(N)$.

%%%%%%%%%%%%%%%
$\bullet SO(4)$
%%%%%%%%%%%%%%%

For the breaking pattern $SO(4)\to  U(2)$,  
by plugging the values $N=4, N_0=0$, and $N_1=2$ into the 
general formula (\ref{formula2}),
one gets $K=\frac{4\eta F}{\La^2}$ and $m=\frac{\eta F}{\La^2}-\eta \La^2$.
As studied in \cite{afo1}, the solutions for the factorization 
problem of degenerated case can be represented as
$
\widetilde{y_m}^2=\left(x^2+D \right)^2+4G $ 
with $ D=\frac{G}{\ep \La^2}-\ep \La^2
$
where $\ep$ is $2$-nd root of unity (Note that $D=b/2$ and $4G=c-b^2/4$
in the notation of \cite{afo1}).
Therefore by identifying $D, G, \ep$ with $m,F, \eta$ respectively,
the two approaches, glueball approach and strong-coupling approach 
are equivalent to each other. 
The special point comes from the 
condition $G=-\La^{4}$ in which the $SO(4)$ goes to $SO(4)$. 
If we define $\rho=-\eta$, we can rewrite the matrix model curve 
as $\widetilde{y_m}^2= \left(x^2+2\rho \La^2 \right)^2-4 \La^4$, 
which agrees 
with the general formula in the equation 
(\ref{sep3general}). 

%since the $K$ becomes ??? $N_0-2=0 or ?1$. 

%%%%%%%%%%%%%%%
$\bullet SO(5)$
%%%%%%%%%%%%%%%

For the breaking pattern $SO(5)\to SO(3)\times U(1)$ 
since the $K$ becomes $\frac{4 \La^6}{F}$, we can write the $m$ as
$
m=-\frac{F^2}{\La^6} + \frac{\La^6}{F}
$.
By identifying $G$ in the result \cite{afo1} with $F$, we 
find an exact agreement.
From the results in \cite{afo1}, the matrix model 
curve is given as
$
\widetilde{y_m}^2=\left(x^2-D \right)^2+4G$ where
$ D=\frac{G^2}{\La^6}-\frac{\La^6}{G}
$.
Note that $G=b/4$ and $D=a$ in the notation of \cite{afo1}.
The particular point comes from the condition $G^3=-\La^{12}$
where $SO(5) \to SO(5)$. 
If we define $\rho=-\eta^2$ where $\eta^3=-1$ (therefore $\rho^3=-1$), 
we can write $G=\eta \La^4=-\rho^2 \La^4$ and then $\widetilde{y_m}^2=
\left(x^2+2\rho \La^2 \right)^2-4\rho^2 \La^4$. 
This curve agrees with the general 
formula (\ref{sep3general}).

%%%%%%%%%%%%%%%%
$\bullet SO(6)$
%%%%%%%%%%%%%%%%

For the breaking pattern $SO(6) \to SO(4)\times U(1)$, since 
the $K$ becomes $\frac{4\eta \La^4}{\sqrt{F}}$, where $\eta^4=1$ 
we can write down 
$
m=\eta \left( \frac{\La^4}{\sqrt{F}}-
\frac{F\sqrt{F}}{\eta^2 \La^4} \right)
$.
By identifying $\epsilon$, and $G^2$ in \cite{afo1} 
with $-\eta$ and $F$ respectively, 
we find an exact agreement between the 
glueball approach and strong-coupling approach.
In \cite{afo1} 
the special point is given by the condition $G^4=\La^8$ where
there exists a breaking pattern $SO(6) \to SO(6)$. Putting this 
value, we obtain the matrix model curve as
$
\widetilde{y_m}^2= \left(x^2+2\rho \La^2 \right)^2 -4 \rho^2 \La^4
$
where $\rho$ is $4$-th root of unity. This curve agrees with the 
one obtained from glueball approach exactly.

%%%%%%%%%%%%%%%%
$\bullet SO(7)$
%%%%%%%%%%%%%%%

For the breaking pattern $SO(7) \to SO(3)\times U(2)$ since 
the $K$ becomes $-\frac{4 \La^{10}}{F^2}$, we can write down
$
m=-\frac{\La^{10}}{F^2}+\frac{F^3}{\La^{10}} 
$.
By identifying $G$ in \cite{afo1} with $F$, we 
find an exact agreement.
In \cite{afo1}, the particular  
point is given by the condition $G^5=-\La^{20}$ where
$SO(7) \to SO(7)$. 
Putting this value, we obtain the matrix model curve
$
\widetilde{y_m}^2=\left(x^2+2\rho  \La^2 \right)^2 -4\rho^2 \La^4
$
where $\rho$ satisfies $\rho^5=-1$. This curve agrees with the one 
obtained from the glueball approach.

When we increase the $N$ beyond 7, we do not have any results for 
the matrix model curve from the strong-coupling approach so far, we have to
resort to the expression (\ref{formula2}) only. In this formula, the 
matrix model curve is given by $N_0, N_1$, and $F$. In general, the $N_0$ and
$N_1$'s are obtained from the relations (\ref{n0n1}). However, contrary to
the nondegenerated case, the contour integrals do not lead to the final 
results for $N_0$ and $N_1$ easily, 
because the matrix model curve does not have 
double root, which will make the $T(x)$ complicated.

%%%%%%%%%%%%%%%%%%%%%%%%%%%%%%%%%%%%%%%%%%%%%%%%%%%%%%%%%%%%%%%%%%%%%%%%%%%%%
%\section{Discussion}
%\setcounter{equation}{0}
%%%%%%%%%%%%%%%%%%%%%%%%%%%%%%%%%%%%%%%%%%%%%%%%%%%%%%%%%%%%%%%%%%%%%%%%%%%%%

%\indent

%Comments on the 1) the $n=0$ and $n=2$ singularity, 2)
%the $n=1$ and $n=2$ singularity, and 3) the 
%$n=0$, $n=1$ and $n=2$ singularity

%%%%%%%%%%%%%%%%%%%%%%%%%%%%%%%%%%%%%%%%%%%%%%%%%%%%%%%%%%%%%%%%%%%%%%%%%%%%%
%\newpage
\vspace{1cm}
\centerline{\bf Acknowledgments}

This research of CA was supported by Kyungpook National University 
Research Team Fund, 2003. 
We thank D. Shih for the correspondence on his paper.
We are grateful to K. Ito, K. Konishi and N. Sakai
for relevant discussions. 
CA thanks Dept. Of Physics, Tokyo Institute of Technology
where part of this work was undertaken.

%%%%%%%%%%%%%%%%%%%%%%%%%%%%%%%%%%%%%%%%%%%%%%%%%%%%%%%%%%%%%%%%%%%%%%%%%%%%%
\appendix

\renewcommand{\thesection}{\large \bf \mbox{Appendix~}\Alph{section}}
\renewcommand{\theequation}{\Alph{section}\mbox{.}\arabic{equation}}

%%%%%%%%%%%%%%%%%%%%%%%%%%%%%%%%%%%%%%%%%%%%%%%%%%%%%%%%%%%%%%%%%%%%%%%%%%%%%
\section{\large \bf The derivatives of dual periods on the $n=1$ branch  }
\setcounter{equation}{0}
%%%%%%%%%%%%%%%%%%%%%%%%%%%%%%%%%%%%%%%%%%%%%%%%%%%%%%%%%%%%%%%%%%%%%%%%%%%%%

\indent

The derivative of $\Pi_0$ with respect to $f_0$
when $x_1$ is arbitrary (not near
the singularity) can be obtained from the definition of $\Pi_0$
in section 2 as follows:
\begin{eqnarray}
4\pi i \frac{\partial \Pi_0}{\partial f_0}&=&\int_0^{\Lambda_0}
\frac{dx}{\sqrt{\left(x^2+x_0^2\right)\left(x^2+x_1^2\right)
\left(x^2+x_2^2 \right)}}=
\frac{1}{2}\int_{0}^{\Lambda_0^2}\frac{dt}{\sqrt{t\left(t+x_0^2\right)
\left(t+x_1^2\right)\left(t+x_2^2\right)}} \nonumber \\
&=&\frac{1}{x_1\sqrt{x_2^2-x_0^2}}F \left(\phi | R \right)
\end{eqnarray}
where we make a change of variable $t=x^2$
and in the final relation we used 
$t= \frac{-y^2}{1-y^2}$
and then $y^2 =\frac{x_2^2}{x_2^2-x_0^2} z^2$ with $\La_0$ large.
Although the dual period $\Pi_0$ is different from the one in
$U(N)$ case, the derivative of $\Pi_0$ with respect to $f_0$ in the
$t$-integration is the same as $\pa \Pi_2 /\pa f_0$ in (A.4) of
\cite{Shih} up to an overall constant. Also we have $x_0 < x_1 < x_2$
as before.
Here  
$\phi, R$ and the first kind elliptic integral $F(\phi | R )$ are
\begin{eqnarray}
\phi & = & \sin^{-1}\sqrt{\frac{x_2^2-x_0^2}{x_2^2}},\qquad 
R=\sqrt{\frac{x_2^2\left(x_1^2-x_0^2\right)}{x_1^2\left(x_2^2-
x_0^2\right)}},\nonu \\
F(\phi| R) & = & \int_0^{\phi}\frac{d \theta}{\sqrt{1-R^2\sin^2 \theta}}
=  \int_0^{w}\frac{d z}{\sqrt{\left(1-z^2 \right) \left(1-R^2 z^2\right)}} 
\equiv
F\left(w=\sin^{-1} \phi | R\right).
\label{elliptic}
\end{eqnarray}

Similarly
the derivative of $\Pi_1$ with respect to $f_0$
is given by 
\begin{eqnarray}
4\pi i \frac{\partial \Pi_1}{\partial f_{0}}&=&\int_{ix_2}^{i 
\Lambda_0}\frac{dx}{\sqrt{\left(x^2+x_0^2\right)\left(x^2+x_1^2\right)
\left(x^2+x_2^2\right)}}=
-\frac{1}{2}\int_{x_2^2}^{\Lambda_0^2}\frac{dt}{\sqrt{t\left(t-x_0^2\right)
\left(t-x_1^2\right)\left(t-x_2^2\right)}} \nonumber  \\
&=&-\frac{1}{x_1\sqrt{x_2^2-x_0^2}}F\left(\psi |R \right)
\end{eqnarray}
where we introduce a new angle $\psi$ as follows: 
\begin{eqnarray}
\psi=\sin^{-1} \left(\frac{x_1}{x_2} \right). 
\end{eqnarray}
From these general formulas we can derive the equations (\ref{del00}) and 
(\ref{del10}) by using the properties of trigonometric functions. 
That is, 
$\phi$ can be represented as
\begin{eqnarray}
\phi=\frac{1}{2}\left[\sin^{-1} \left( \frac{x_2^2-2x_0^2}{x_2^2}
\right)+
\frac{\pi}{2} \right]. 
\end{eqnarray}
Under the limit, $x_1\to x_0$ (at the singularity), the  
parameter $R$ goes to zero and the elliptic 
function behaves as $F(\phi |0)=\phi$ from the (\ref{elliptic}). 
After some calculations, 
we 
obtain the following 
relations which are the same as (\ref{del00}) and (\ref{del10}) 
precisely,
\begin{eqnarray}
4\pi i \frac{\partial \Pi_0}{\partial f_0}&=&\frac{1}{2x_0 
\sqrt{x_2^2-x_0^2}}\left[\sin^{-1} \left(
\frac{x_2^2-2x_0^2}{x_2^2} \right)+
\frac{\pi}{2} \right], \nonumber \\
4\pi i \frac{\partial \Pi_1}{\partial f_0}&=&\frac{-1}{2x_0 
\sqrt{x_2^2-x_0^2}}\left[-\sin^{-1} \left(
\frac{x_2^2-2x_0^2}{x_2^2} \right)+
\frac{\pi}{2} \right],
\end{eqnarray}
where we use the fact that $\psi=\phi+\pi/2$ under this limit.

The derivative of $\Pi_0$ with respect to $f_2$ (when we consider 
the type of integration, this corresponds to $\pa \Pi_2/\pa f_1$
of $U(N)$ case)
can be obtained from the definition of $\Pi_0$
by the same change of variables before  
\begin{eqnarray}
4\pi i \frac{\partial \Pi_0}{\partial f_2}&=&\frac{1}{2}
\int_0^{\Lambda^2_0}\frac{t \ dt}{\sqrt{t\left(t+x_0^2\right)
\left(t+x_1^2\right)\left(t+x_2^2\right)}} \nonu  \\
&=& \frac{x_0^2}{x_1\sqrt{x_2^2-x_0^2}}\int_0^{\sqrt{
\frac{\Lambda_0^2}{n^2\left(\Lambda_0^2+x_0^2\right)}}}
\left( \frac{1}{1-n^2 z^2}
-1 \right) 
\frac{dz}{\sqrt{\left(1-z^2\right)\left(1-R^2z^2\right)}} \nonu \\
&=& \frac{x_0^2}{x_1\sqrt{x_2^2-x_0^2}} \left[\Pi 
\left(n\ ;\ \sin^{-1} \left[
\frac{1}{n}\sqrt{\frac{\Lambda_0^2}{\Lambda_0^2+x_0^2}} \right] 
\Bigg| R 
\right)-F\left(\phi | R\right) 
\right] \nonu \\
&= & \frac{x_0^2}{x_1\sqrt{x_2^2-x_0^2}} \Pi 
\left(n\ ;\ \sin^{-1} \left[
\frac{1}{n}\sqrt{\frac{\Lambda_0^2}{\Lambda_0^2+x_0^2}} \right] 
\Bigg| R 
\right) - 4\pi i x_0^2 \frac{\pa \Pi_0}{\pa f_0} 
\label{del0del2825}
\end{eqnarray}
where we introduce a new notation for $\phi$  and the third 
kind elliptic integral $\Pi(c; \phi|R)$
is given by 
\begin{eqnarray}
n^2 & \equiv & \frac{x_2^2}{x_2^2-x_0^2}=\frac{1}{\sin \phi},
\nonu \\
\Pi(c; \phi|R)
 & = & \int_0^{\phi}\frac{d \theta}{\left(1-c^2 \sin^2 \th \right) 
\sqrt{1-R^2\sin^2 \theta}} \nonu \\
& = & \int_0^w \frac{dz}{\left(1-c^2 z^2\right)\sqrt{\left(1-z^2\right)
\left(1-R^2 z^2\right)}}
\equiv 
\Pi\left(c; w=\sin^{-1} \phi|R \right).
\end{eqnarray}
%where we use a notation for $\Pi$ which is not standard one. 
Note that if we take a limit $\Lambda_0^2\to \infty$,  the first term of
(\ref{del0del2825}) contains a
divergence term. Therefore, we should keep $\Lambda_0$ term 
in the formula explicitly.

Next we evaluate this formula at the singular point with $x_1^2 \to 
x_0^2$. Under this limit, since the $R$ goes to zero we have only to 
consider the following integral,
\begin{eqnarray}
\int_{0}^{\frac{1}{n}+\epsilon}\frac{dz}{\left(1-n^2z^2\right) 
\sqrt{\left(1-z^2\right)}} & =& \frac{1}{\sqrt{n^2-1}} \tanh^{-1}
\left( \frac{\sqrt{n^2-1}z}{\sqrt{1-z^2}}\right) 
\Bigg|_{0}^{\frac{1}{n}+\ep} \nonu \\
&\simeq &\frac{-1}{2\sqrt{n^2-1}} \log \Bigg|  \frac{\ep n^3}{
2\left(n^2-1 \right)} 
 \Bigg| \nonu \\
&=& \frac{-1}{2\sqrt{n^2-1}} \log 
\Bigg| \frac{x_2^2}{4\Lambda_0^2} \Bigg|
\label{expre}
\end{eqnarray}
where we define 
\bea
\epsilon \equiv -\frac{x_0^2}{2n \Lambda_0^2}
\eea
and 
drop out $\mathcal{O}(\epsilon)$ in (\ref{expre}).
We also used the fact that $\tanh^{-1} x = \frac{1}{2} \log \Bigg| 
\frac{1+x}{1-x} \Bigg|$.
Taking into account an overall factor and $x_1^2\to x_0^2$, the 
first equation of (\ref{del0del2825}) becomes 
\begin{eqnarray}
-\frac{1}{2}\log \Bigg | \frac{x_2^2}{4\Lambda_0^2} \Bigg |. 
\label{fin825}
\end{eqnarray}
Therefore (\ref{del0del2825}) reproduces to (\ref{02der}) at the
singularity.

Similarly one can execute the integral
\begin{eqnarray}
4\pi i \frac{\partial \Pi_1}{\partial f_2}&=&\frac{1}{2}
\int_{x_2^2}^{\Lambda^2_0}\frac{t \ dt}{\sqrt{t\left(t-x_0^2\right)
\left(t-x_1^2\right)
\left(t-x_2^2\right)}} \nonu  \\
&=& \frac{\left(x_2^2-x_1^2\right)}{x_1\sqrt{x_2^2-x_0^2}}\int_{0}^
{\sqrt{\frac{x_1^2 \left(\Lambda_0^2-x_2^2 \right)}
{x_2^2\left(\Lambda_0^2-x_1^2\right)}}}
\left( \frac{1}{1-\tilde{n}^2z^2} +
\frac{x_1^2}{x_2^2-x_1^2} \right) 
\frac{dz}{\sqrt{\left(1-z^2\right)\left(1-R^2z^2\right)}} \nonu \\
&=& \frac{\left(x_2^2-x_1^2\right)}{x_1\sqrt{x_2^2-x_0^2}} \left[ \Pi 
\left( \tilde{n} \ ; \ \sin^{-1} \left[
\sqrt{\frac{\Lambda_0^2-x_2^2}{\Lambda_0^2-x_1^2}} 
\sin \psi \right] \Bigg|R \right)+ 
\frac{x_1^2}{\left(x_2^2-x_1^2\right)}F 
\left(\psi |R \right) \right] \nonu \\
& = & \frac{\left(x_2^2-x_1^2\right)}{x_1\sqrt{x_2^2-x_0^2}} \Pi 
\left( \tilde{n} \ ; \ \sin^{-1} \left[
\sqrt{\frac{\Lambda_0^2-x_2^2}{\Lambda_0^2-x_1^2}} 
\sin \psi \right] \Bigg|R \right) 
-4\pi i x_1^2 \frac{\pa \Pi_1}{\pa f_0}
\label{del1del2825}
\end{eqnarray}
where 
\bea
\tilde{n}^2 \equiv \frac{x_2^2}{x_1^2} = \frac{1}{\sin^2 \psi}, \qquad
\tilde{\ep} \equiv -\frac{1}{2} \frac{x_0\left(x_0^2-x_2^2 \right)}
{x_2\left(\La_0^2-x_0^2 \right)}.
\eea
As in previous 
calculation, we can evaluate the first term of (\ref{del1del2825}) 
at the singularity and 
obtain the same result (\ref{fin825}). Therefore, we reproduced 
the relation (\ref{12der}).

%%%%%%%%%%%%%%%%%%%%%%%%%%%%%%%%%%%%%%%%%%%%%%%%%%%%%%%%%%%%%%%%%%%%%%%%%%%%%
\section{\large \bf The derivatives of periods on the $n=1$ branch  }
\setcounter{equation}{0}
%%%%%%%%%%%%%%%%%%%%%%%%%%%%%%%%%%%%%%%%%%%%%%%%%%%%%%%%%%%%%%%%%%%%%%%%%%%%%

\indent

Next let us consider the derivatives of $S_1$. 
By changing the variable $t$ to $z$,
\begin{eqnarray}
t \equiv \frac{x_1^2-x_0^2y^2}{1-y^2},
\qquad y^2 \equiv \frac{x_2^2-x_1^2}{x_2^2-x_0^2}z^2
\end{eqnarray}
we can get the following formula:
\begin{eqnarray}
4\pi i \frac{\partial S_1}{\partial f_0}&=&\int_{ix_1}^{ix_2}\frac{dx}
{\sqrt{\left(x^2+x_0^2\right)\left(x^2+x_1^2\right)\left(x^2+x_2^2\right)}}
=\int_{x_1^2}^{x_2^2}
\frac{dt}{2\sqrt{t\left(t-x_0^2\right)\left(t-x_1^2\right)
\left(t-x_2^2\right)}} \nonu \\
&=&\frac{i}{x_1\sqrt{x_2^2-x_0^2}}F\left( \frac{\pi}{2} \Bigg| 
k^{\prime} \right)
\end{eqnarray}
where we introduce
\begin{eqnarray}
n^{\prime}=\frac{x_2^2-x_1^2}{x_2^2-x_0^2},\qquad k^{\prime ^2}=
\frac{x_0^2\left(x_2^2-x_1^2\right)}{x_1^2\left(x_2^2-x_0^2\right)}.
\label{def}
\end{eqnarray}
Although the form of $S_1$ is different from the one in $U(N)$ case, 
the above $t$-integration is the same as $\pa S_2/\pa f_0$ in (A.8)
of \cite{Shih}. 
Moreover one has 
\begin{eqnarray}
4\pi i \frac{\partial S_1}{\partial f_2}&=&\int_{ix_1}^{ix_2}
\frac{x^2dx}{\sqrt{\left(x^2+x_0^2\right)\left(x^2+x_1^2\right)
\left(x^2+x_2^2\right)}}=\int_
{x_1^2}^{x_2^2}\frac{t \ dt}{2\sqrt{t\left(t-x_0^2\right)
\left(t-x_1^2\right)\left(t-x_2^2\right)}} 
\nonu \\
&=&\frac{i\left(x_1^2-x_0^2\right)}{x_1\sqrt{x_2^2-x_0^2}}\Pi
 \left(n^{\prime} \  ; \ \frac{\pi}{2} \Bigg| k^{\prime} \right)+\frac{ix_0^2}
{x_1\sqrt{x_2^2-x_0^2}}F \left( \frac{\pi}{2} \Bigg| k^{\prime} \right).
\end{eqnarray}
This corresponds to the $\pa S_2/ \pa f_1$ for $U(N)$ case.
We can evaluate the above formula under the limit 
$x_1^2\to x_0^2$ and in this case $k^{\prime} \rightarrow 1$:
\begin{eqnarray}
F \left(\frac{\pi}{2} \Bigg|k^{\prime} \right) & \to & \frac{1}{2}\log 
\left(\frac{16}{1-k^{\prime 2}} \right)+ {\cal O}
\left(1-k^{\prime 2}\right), \nonu \\
\Pi \left(n^{\prime} \ ;  \ \frac{\pi}{2} \Bigg| k^{\prime} \right)
& \to & 
\frac{\left(x_2^2-x_0^2 \right)}{\left(x_1^2-x_0^2
\right)}\sqrt{\frac{x_1^2}{x_2^2-x_1^2}}
\sin ^{-1}
\left(\sqrt{\frac{x_2^2-x_0^2}{x_2^2}} \right).
\end{eqnarray}
These relations were given already in \cite{Shih}.

Let us compute the partial derivative $\frac{\pa \Pi_1 }{\pa S_0 }$
near the $n=0$ and $n=1$ singularity that is necessary to 
the coupling constant at the singularity.
By using the chain rule one writes   
\begin{eqnarray}
\frac{\partial \Pi_1}{\partial S_0}=\frac{\partial \Pi_1}
{\partial f_2}\frac{\partial f_2}{\partial S_0}+
\frac{\partial \Pi_1}{\partial f_0}\frac{\partial f_0}
{\partial S_0}=\frac{1}{4} \left(\kappa \frac{\partial \Pi_1}{\pa f_0}-
\frac{\partial \Pi_1}{\partial f_2} \right)
\end{eqnarray}
where we use the relation $S_0 = S-2S_1=-4f_2-2S_1$ in order to
eliminate $S_0$ and  define the $\kappa$ as follows:
\begin{eqnarray}
\kappa \equiv \frac{\frac{\partial S_1}{\partial f_2}}{
\frac{\partial S_1}{\partial f_0}}=x_0^2+\frac{2
\sqrt{x_0^2\left(x_2^2-x_0^2\right)}}{\log \left(\frac{16}{1-
k^{\prime 2}}\right)}\sin ^{-1}
\left( \sqrt{\frac{x_2^2-x_0^2}{x_2^2}} \right).
\end{eqnarray}
Inserting the results for $x_2$ and $x_0$ given in
(\ref{sol}), we can rewrite as 
\begin{eqnarray}
\frac{\partial \Pi_1}{\partial S_0} & \simeq &
\frac{1}{4} \times \frac{2
\sqrt{x_0^2\left(x_2^2-x_0^2\right)}}{\log \left(\frac{16}{1-
k^{\prime 2}}\right)}\sin ^{-1}
\left( \sqrt{\frac{x_2^2-x_0^2}{x_2^2}} \right) 
\times  \frac{\partial \Pi_1}{\pa f_0}
\nonu \\
&= &
  \frac{i \pi N_1 \left(N_0-2 \right)}{16 \left(N-2 \right)^2} 
\frac{1}{ \log \left(\frac{16}{1-k^{\prime 2}}\right)}
\label{coupling}
\end{eqnarray}
where we used 
$
\frac{\partial \Pi_1}{\pa f_0}=\frac{i}{8  } \frac{1}{x_0\sqrt{
x_2^2-x_0^2}} \left(1 -\frac{2N_1}{N-2} \right)
$. For $USp(2N)$ gauge theory, the computation can be done similarly.

\end{document}